\documentclass[12pt]{article}
\usepackage{enumerate}
\usepackage{amsfonts}
\usepackage{amsmath}
\usepackage{amssymb}
\usepackage{amsthm}
\usepackage{latexsym}
\usepackage{color}

\def\A{\mathfrak{A}}
\def\D{\mathcal{D}}
\def\H{\mathcal{H}}

\def\M{\mathbb{M}}

\def\S{\mathfrak{S}}
\def\C{\mathfrak{C}}

\def\T{\mathfrak{T}}

\newcommand{\supp}{\mathrm{supp}}
\newcommand{\rank}{\mathrm{rank}}
\newcommand{\id}{\mathrm{Id}}
\newcommand{\Tr}{\mathrm{Tr}}

\newcommand{\shs}{\hspace{1pt}}
\newcounter{defin}  \newcounter{lemma}  \newcounter{theorem}
\newcounter{property} \newcounter{corol}  \newcounter{remark} \newcounter{example}
\newenvironment{lemma}{\par\refstepcounter{lemma}     \textbf{Lemma \thelemma.} }{\rm\par}
\newenvironment{theorem}{\par\refstepcounter{theorem}     \textbf{Theorem \thetheorem.}\ }{\rm\par}

\newenvironment{corollary}{\par\refstepcounter{corol}     \textbf{Corollary \thecorol.} }{\rm\par}

\begin{document}

%\title{Convergence conditions for the quantum relative entropy and other applications of the deneralized Dini-type lemma}

\title{Convergence conditions for the quantum relative entropy and other applications of the deneralized quantum Dini lemma}

%\title{The quantum relative entropy: the criterion of convergence and the preservation of convergence under the action of CP linear maps}

\author{M.E. Shirokov\footnote{email:msh@mi.ras.ru}\\Steklov Mathematical Institute, Moscow}
\date{}
\maketitle

%UDC: 519.248.3

%MSC: 81P45, 94A17, 46L53

\begin{abstract}
 We describe a generalized version of the result called  quantum Dini lemma that was used previously for
analysis of local continuity of basic correlation and entanglement measures. The generalization consists in considering sequences of functions instead of a single function. It allows us to expand the scope of possible applications of the method. We prove
two general dominated convergence theorems and the theorem about preserving  local continuity under convex mixtures.

By using these theorems we obtain several convergence conditions for the quantum relative entropy and for the mutual information of a quantum channel considered as a function
of a pair (channel, input state). A simple convergence criterion for the von Neumann entropy is also obtained.
\end{abstract}

\section{Introduction}

The rigorous mathematical study of quantum systems and channels of infinite dimension requires considerable efforts due to the fact that the basic characteristics of such systems and channels have singular properties (discontinuity, infinite or indefinite values, etc). One of the main ways  of studying such characteristics is the approximation method, which allows us to transfer to the infinite-dimensional case the results proved in finite-dimensional settings, prove uniform continuity bounds and other estimates, etc.

In \cite[Section 4.2]{QC} the specific approximative method for studying the characteristics of infinite-dimensional quantum systems
based on the general result called quantum Dini lemma was proposed. This result allows us to obtain uniform approximation of
nonnegative lower semicontinuous functions satisfying  the inequality
\begin{equation}\label{LAA-1+}
  f(p\rho+(1-p)\sigma)\geq pf(\rho)+(1-p)f(\sigma)-a_f(p)
\end{equation}
for any $\rho$ and $\sigma$ in $\S(\H)$ and $p\in(0,1)$, where $a_f(p)$ is a nonnegative continuous function vanishing as $\,p\to0$, that can be  treated
as a weakened form of concavity (a real concavity corresponds to the case $a_f(p)\equiv0$). These properties are fulfilled for many important characteristics,
basic examples are the following:
\begin{itemize}
   \item the von Neumann entropy $S(\rho)$, the output entropy $S(\Phi(\rho))$ and  the entropy exchange $S(\Phi,\rho)$ of a quantum channel $\Phi$ \cite{H-SCI,L-2,W};
   \item the quantum relative entropy $D(\rho\shs\|\shs\sigma)$ as a function of $\rho$ \cite{H-SCI,L-2,W};
   \item the quantum conditional mutual information $I(A\!:\!B|C)_{\rho}$ \cite[Section 5.2]{QC};
   \item the relative entropy distance to a set of quantum states satisfying the particular regularity condition (see \cite{L&Sh}),
   in particular, the relative entropy of entanglement.
   \item the constrained Holevo capacity and the mutual information of a quantum channel (as a function of input state) \cite{H-SCI,Wilde-new},\cite[Section 5.5]{QC}.
\end{itemize}

In \cite{QC} the quantum Dini lemma and its corollaries were used to obtain local continuity conditions
for basic correlation and entanglement measures in composite quantum systems (quantum conditional mutual information,
one-way classical correlation, quantum discord, entanglement of formation and its regularization).

In this article we prove a generalized version of quantum Dini lemma which allows us to analyse sequences
of nonnegative lower semicontinuous functions satisfying  inequality (\ref{LAA-1+}). This version is applied to prove
two general dominated convergence theorems and the theorem about preserving  local continuity under convex mixtures.

These theorems are used to obtain convergence conditions for the quantum relative entropy $D(\rho\shs\|\shs\sigma)$ and for the mutual information $I(\Phi,\rho)$ of a quantum channel $\Phi$ at a state $\rho$ considered as a function
of a pair (channel, input state), i.e. conditions under which the limit relations
$$
\lim_{n\to+\infty}D(\rho_n\|\shs\sigma_n)=D(\rho_0\|\shs\sigma_0)<+\infty,\qquad
\lim_{n\to+\infty}I(\Phi_n,\rho_n)=I(\Phi_0,\rho_0)<+\infty
$$
hold for given sequences $\{\rho_n\}$ and $\{\sigma_n\}$ of quantum states converging to states $\rho_0$ and $\sigma_0$  and a sequence
$\{\Phi_n\}$ of quantum channels strongly converging to a channel $\Phi_0$.

\section{Preliminaries}

Let $\mathcal{H}$ be a separable Hilbert space,
$\mathfrak{B}(\mathcal{H})$ the algebra of all bounded operators on $\mathcal{H}$ with the operator norm $\|\cdot\|$ and $\mathfrak{T}( \mathcal{H})$ the
Banach space of all trace-class
operators on $\mathcal{H}$  with the trace norm $\|\!\cdot\!\|_1$. Let
$\mathfrak{S}(\mathcal{H})$ be  the set of quantum states (positive operators
in $\mathfrak{T}(\mathcal{H})$ with unit trace) \cite{H-SCI,BSimon,Wilde}.

Denote by $I_{\mathcal{H}}$ the unit operator on a Hilbert space
$\mathcal{H}$ and by $\id_{\mathcal{\H}}$ the identity
transformation of the Banach space $\mathfrak{T}(\mathcal{H})$.

The \emph{von Neumann entropy} of a quantum state
$\rho \in \mathfrak{S}(\H)$ is  defined by the formula
$S(\rho)=\Tr\eta(\rho)$, where  $\eta(x)=-x\ln x$ if $x>0$
and $\eta(0)=0$. It is a concave lower semicontinuous function on the set~$\mathfrak{S}(\H)$ taking values in~$[0,+\infty]$ \cite{H-SCI,L-2,W}.
The von Neumann entropy satisfies the inequality
\begin{equation}\label{w-k-ineq}
S(p\rho+(1-p)\sigma)\leq pS(\rho)+(1-p)S(\sigma)+h_2(p)
\end{equation}
valid for any states  $\rho$ and $\sigma$ in $\S(\H)$ and $p\in(0,1)$, where $\,h_2(p)=\eta(p)+\eta(1-p)\,$ is the binary entropy \cite{N&Ch,Wilde}.

We will use the  homogeneous extension of the von Neumann entropy to the positive cone $\T_+(\H)$ defined as
\begin{equation}\label{S-ext}
S(\rho)\doteq(\Tr\rho)S(\rho/\Tr\rho)=\Tr\eta(\rho)-\eta(\Tr\rho)
\end{equation}
for any nonzero operator $\rho$ in $\T_+(\H)$ and equal to $0$ at the zero operator \cite{L-2}.

By using concavity of the entropy and inequality (\ref{w-k-ineq}) it is easy to show that
\begin{equation}\label{w-k-ineq+}
S(\rho)+S(\sigma)\leq S(\rho+\sigma)\leq S(\rho)+S(\sigma)+H(\{\Tr\rho,\Tr\sigma\}),
\end{equation}
for any $\rho$ and $\sigma$ in $\T_+(\H)$, where $H(\{\Tr\rho,\Tr\sigma\})=\eta(\Tr\rho)+\eta(\Tr\sigma)-\eta(\Tr(\rho+\sigma))$
is the homogeneous extension of the binary entropy to the positive cone in $\mathbb{R}^2$.

The \emph{quantum relative entropy} for two states $\rho$ and
$\sigma$ in $\mathfrak{S}(\mathcal{H})$ is defined as
\begin{equation}\label{qre-def}
D(\rho\shs\|\shs\sigma)=\sum_i\langle
i|\,\rho\ln\rho-\rho\ln\sigma\,|i\rangle,
\end{equation}
where $\{|i\rangle\}$ is the orthonormal basis of
eigenvectors of the state $\rho$ and it is assumed that
$D(\rho\,\|\sigma)=+\infty$ if $\,\mathrm{supp}\rho\shs$ is not
contained in $\shs\mathrm{supp}\shs\sigma$ \cite{H-SCI,L-2,Wilde}.\footnote{The support $\mathrm{supp}\rho$ of a state $\rho$ is the closed subspace spanned by the eigenvectors of $\rho$ corresponding to its positive eigenvalues.}

The \emph{quantum mutual information} (QMI) of a state $\,\rho\,$ in $\S(\H_{AB})$  is defined as
\begin{equation}\label{mi-d}
I(A\!:\!B)_{\rho}=D(\rho\shs\Vert\shs\rho_{A}\otimes
\rho_{\shs B})=S(\rho_{A})+S(\rho_{\shs B})-S(\rho),
\end{equation}
where the second formula is valid if $\,S(\rho)\,$ is finite \cite{L-mi}.
Basic properties of the relative entropy show that $\,\rho\mapsto
I(A\!:\!B)_{\rho}\,$ is a lower semicontinuous function on the set
$\S(\H_{AB})$ taking values in $[0,+\infty]$.\smallskip

\textbf{Remark 1.} The function $\,\rho\mapsto
I(A\!:\!B)_{\rho}\,$ is uniformly continuous on the set $\S(\H_{AB})$ if (and only if) $\min\{\dim\H_A,\dim\H_B\}<+\infty$  \cite[Section 5.2.1]{QC}.
\smallskip

It is well known that
\begin{equation}\label{MI-UB}
I(A\!:\!B)_{\rho}\leq 2\min\{S(\rho_A),S(\rho_B)\}
\end{equation}
for any state $\rho\in\S(\H_{AB})$, where $\rho_A\doteq\Tr_B\rho$ and $\rho_B\doteq\Tr_A\rho$ \cite{L-mi,H-SCI}.
\smallskip

A \emph{quantum channel} $\Phi$ from a system $A$ to a system $B$ is a completely positive trace preserving
linear  map from $\T(\H_A)$ to $\T(\H_B)$  \cite{H-SCI,N&Ch,Wilde}.\smallskip

Let $H$ be a positive (semi-definite) operator on a Hilbert space $\mathcal{H}$ with the domain $\mathcal{D}(H)$.  We will assume that
\begin{equation}\label{H-as}
\mathrm{Tr} H\rho=
\left\{\begin{array}{l}
        \sup_n\mathrm{Tr} P_nH\rho\;\; \textrm{if}\;\;  \supp\rho\subseteq {\rm cl}(\mathcal{D}(H))\\
        +\infty\;\;\textrm{otherwise}
        \end{array}\right.
\end{equation}
for any positive operator $\rho\in\T(\H)$, where $P_n$ is the spectral projector of $H$ corresponding to the interval $[0,n]$ and ${\rm cl}(\mathcal{D}(H))$
is the closure of $\mathcal{D}(H)$.

In this article we will deal with unbounded operators $H$ having  discrete spectrum of finite multiplicity. Such operators can be represented as follows
\begin{equation*}%\label{H-rep}
H=\sum_{k=0}^{+\infty} E_k |\tau_k\rangle\langle\tau_k|,
\end{equation*}
where
$\mathcal{T}=\left\{\tau_k\right\}_{k=0}^{+\infty}$ is the set of orthogonal unit
eigenvectors of $H$ corresponding to the nondecreasing sequence $\left\{\smash{E_k}\right\}_{k=0}^{+\infty}$ of eigenvalues
tending to $+\infty$ and \emph{it is implied that the domain $\D(H)$ of $H$ lies within the subspace  $\H_\mathcal{T}$ denerated by $\mathcal{T}$}.
It is easy to see that
\begin{equation}\label{Tr-exp}
  \Tr H\rho=\sum_{k=0}^{+\infty} E_k\langle\tau_k|\rho|\tau_k\rangle
\end{equation}
for any operator $\rho$ in $\T_+(\H)$ such that $\,\supp \rho\,$ lies within the subspace $\H_\mathcal{T}$.

Following the notation used in \cite{QC} we will say that a function $f$ on a convex subset $\S_0$ of $\S(\H)$
is a \emph{locally almost affine} function (briefly, \emph{LAA function}) if it satisfies the inequalities
\begin{equation}\label{LAA-1}
  f(p\rho+(1-p)\sigma)\geq pf(\rho)+(1-p)f(\sigma)-a_f(p)
\end{equation}
and
\begin{equation}\label{LAA-2}
  f(p\rho+(1-p)\sigma)\leq pf(\rho)+(1-p)f(\sigma)+b_f(p),
\end{equation}
for all states $\rho$ and $\sigma$ in $\S_0$ and any $p\in[0,1]$, where $a_f(p)$ and $b_f(p)$ are vanishing functions as $p\rightarrow+0$ (depending of $f$).
These  inequalities can be treated, respectively, as weakened forms of concavity and convexity. For technical simplicity we will assume that
\begin{equation}\label{a-b-assump}
 \textrm{ the functions}\;\; a_f\;\; \textrm{and}\;\; b_f\;\; \textrm{are non-decreasing on}\;\; \textstyle[0,\frac{1}{2}].
\end{equation}

The class of LAA functions on $\S(\H)$ and on convex subsets of $\S(\H)$ includes many important characteristics of quantum systems and channels (von Neumann entropy,  quantum conditional entropy, quantum relative entropy as a function of the first argument,  etc.) \cite{MCB,QC}.

\pagebreak

\section{Generalized quantum Dini lemma}

In this section we consider a generalized version of the technical results proposed and used
in \cite{QC} for analysis of continuity of characteristics of infinite-dimensional quantum systems.
These results (Propositions 1 and 2 in \cite{QC}) allow us to analyse properties of a given
function $f$ on a convex set $\S_0$ of quantum states satisfying either both inequalities
(\ref{LAA-1}) and (\ref{LAA-2}) or one of these inequalities. Below we will prove a
generalized version of Proposition 2 in \cite{QC} that gives possibility to analyse sequences of functions
possessing these properties.

Let $\A$ be a subset of $\S(\H)$ and $\{\Psi_m\}_{m\in\mathbb{N}}$ a family of maps from
$\A$ to $\T_{+}(\H)$ such that
\begin{itemize}
  \item $\Psi_m(\rho)\leq\Psi_{m+1}(\rho)\leq\rho\,$ for any $\,\rho\in\A$ and all $\,m$;
  \item $\lim_{m\to+\infty}\Psi_m(\rho)=\rho\,$ for any $\rho\in\A$;
  \item for any state $\rho\in\A$ there is a countable subset  $M_{\rho}$ of $\,\mathbb{N}\,$ such that
  $\,\lim\limits_{n\to+\infty}\Psi_m(\rho_n)=\Psi_m(\rho)$ for any $\,m\in M_{\rho}$ and any sequence $\{\rho_n\}\subset\A$ converging to a state $\rho\in\A$.
\end{itemize}

The properties of the family $\{\Psi_m\}$ allows us to prove the following\smallskip

\begin{lemma}\label{DTL-L}\cite{QC} \emph{If $\,\C$ is a compact subset of $\,\A$ then}
\begin{equation*}%\label{T-u-c}
\lim_{m\to+\infty}\inf_{\rho\in\C}\Tr\Psi_m(\rho)=1.
\end{equation*}
\end{lemma}\smallskip

The examples of families $\{\Psi_m\}$  are described at the end of Section 4.2.1 in \cite{QC}.

Throughout this section we will use \textbf{the following notation}
\begin{equation}\label{BN}
  [\sigma]\doteq \sigma/\Tr\sigma,\;\; \sigma\in\T_+(\H)\setminus\{0\},\quad\textrm{and}\quad[0]=0.
\end{equation}

We consider the state $[\Psi_m(\rho)]$ as an approximation
for $\rho$. If $\S_0$ is a convex subset of $\S(\H)$ that
contains $\A$ and all the states $[\Psi_m(\rho)]$, $\rho\in\A$, and
$f$ is a nonnegative lower semicontinuous function on $\S_0$
satisfying inequality \eqref{LAA-1} then Proposition 1 in \cite[Section 4.2.1]{QC} implies that
\begin{equation}\label{p-w-c}
\lim_{m\to+\infty} f([\Psi_m(\rho)])=f(\rho)\leq+\infty\quad \forall\rho\in\A,
\end{equation}
\begin{equation}\label{p-w-c+}
\lim_{m\to+\infty} (1-\Tr\Psi_m(\rho))f([\rho-\Psi_m(\rho)])=0\quad  \forall\rho\in\A\;\; \textrm{s.t.}\;  f(\rho)<+\infty,
\end{equation}
where it is assumed that $f(0)=0$, and that these  convergences are uniform on any compact subset of $\A$ on which the
function $f$ is continuous. Proposition 2 in \cite[Section 4.2.1]{QC} generalizes this observation. It states, in particular, that
\begin{equation*}%\label{A+}
\begin{array}{rl}
0\,\leq &\!\!\displaystyle\liminf_{m\to+\infty}\,\inf_{n\geq0}\left(f(\rho_n)-f([\Psi_m(\rho_n)])\right)\\\\
\leq & \!\!\displaystyle\limsup_{m\to+\infty}\,\sup_{n\geq0}\left(f(\rho_n)-f([\Psi_m(\rho_n)])\right)\leq  \limsup_{n\to+\infty} f(\rho_n)-f(\rho_0)
\end{array}
\end{equation*}
and
\begin{equation*}%\label{A+}
\limsup_{m\to+\infty}\,\sup_{n\geq0}(1-\Tr\Psi_m(\rho_n))f([\rho_n-\Psi_m(\rho_n)])\leq \limsup_{n\to+\infty} f(\rho_n)-f(\rho_0)
\end{equation*}
for any sequence $\{\rho_n\}\subseteq\A$ converging to a state $\rho_0\in\A$
provided that $f(\rho_0)$ and $\limsup_{n\to+\infty} f(\rho_n)$ are finite.

The following proposition extends this result to sequences of functions.\smallskip

\textbf{Proposition 1.} \emph{Let $\{\rho_n\}$ be a sequence in $\S(\H)$ converging to a state $\rho_0$. Let $\{\Psi_m\}$
be a family of maps from $
\,\A=\{\rho_n\}\cup\{\rho_0\}$ to $\T_+(\H)$ possessing the properties described before Lemma \ref{DTL-L} and $M_{\rho_0}$ the corresponding countable subset of $\,\mathbb{N}$. Let $\S_0$ be a convex subset of $\S(\H)$ containing all the states $\rho_n$ and $[\Psi_m(\rho_n)]$, $n\geq0$.}

\emph{Let $\{f_n\}_{n\geq 0}$ be a sequence of nonnegative lower semicontinuous functions on $\S_0$ satisfying inequality \eqref{LAA-1} with the same function $a_f$ (not depending on $n$) such that $f_n(\rho_n)$ is finite all $n\geq 0$,
\begin{equation}\label{b-cond}
A\doteq\limsup_{n\to+\infty} f_n(\rho_n)-f_0(\rho_0)\in[0,+\infty)\,\quad and \,\quad \liminf_{n\to+\infty} f_n([\Psi_m(\rho_n)])\geq f_0([\Psi_m(\rho_0)])
\end{equation}
for all sufficiently large $\,m\in M_{\rho_0}$. Then
\begin{equation}\label{A+}
\begin{array}{rl}
0\,\leq &\!\!\displaystyle\liminf_{m\to+\infty}\,\inf_{n\geq0}\left(f_n(\rho_n)-f_n([\Psi_m(\rho_n)])\right)\\\\
\leq & \!\!\displaystyle\limsup_{m\to+\infty}\,\sup_{n\geq0}\left(f_n(\rho_n)-f_n([\Psi_m(\rho_n)])\right)\leq A
\end{array}
\end{equation}
and
\begin{equation}\label{B+}
\limsup_{m\to+\infty}\,\sup_{n\geq0} \Tr\Delta_m(\rho_n)f_n([\Delta_m(\rho_n)])\leq A,
\end{equation}
where $[\sigma]$ denotes the state proportional to a positive operator $\sigma$ and $\Delta_m=\id_{\H}-\Psi_m$.}\smallskip

\emph{Let $\{g_n\}_{n\geq 0}$ be a sequence of functions on $\S_0$ such that $|g_n(\rho)|\leq f_n(\rho)$ for any $\rho$ in $\S_0$ and all $\,n\geq0$. If
all the functions $g_n$  satisfy inequality \eqref{LAA-1} with the same function $a_g$ (not depending on $n$) then
\begin{equation}\label{C-1+}
\limsup_{m\to+\infty}\,\sup_{n\geq0}\left(g_n([\Psi_m(\rho_n)])-g_n(\rho_n)\right)\leq A.
\end{equation}
If all the functions $g_n$  satisfy inequality \eqref{LAA-2} with the same function $b_g$ (not depending on $n$) then
\begin{equation}\label{C-2+}
\limsup_{m\to+\infty}\,\sup_{n\geq0}\left(g_n(\rho_n)-g_n([\Psi_m(\rho_n)])\right)\leq A.
\end{equation}
If all the functions $g_n$  satisfy both inequalities \eqref{LAA-1} and  \eqref{LAA-2} with the same functions $a_g$  and $b_g$ (not depending on $n$) then}
\begin{equation*}%\label{C-2+}
\limsup_{m\to+\infty}\,\sup_{n\geq0}\left|g_n(\rho_n)-g_n([\Psi_m(\rho_n)])\right|\leq A.
\end{equation*}

\emph{Proof.} Let $\mu_n^m=\Tr\Psi_m(\rho_n)$. By Lemma \ref{DTL-L} we have
\begin{equation}\label{mu-u-c}
\lim_{m\to+\infty}\inf_{n\geq 0}\mu_n^m=1.
\end{equation}

Let $\varepsilon\in(0,1/2)$ be arbitrary. It follows from \eqref{p-w-c} and \eqref{mu-u-c} that
there is $m_\varepsilon\in M_{\rho_0}$ such that
\begin{equation}\label{m-e}
|f_0(\rho_0)-f_0([\Psi_{m_\varepsilon}(\rho_0)])|<\varepsilon,\quad \mu_n^{m_\varepsilon}>1-\varepsilon,\quad   a_f(1-\mu_n^{m_\varepsilon})<\varepsilon\quad \forall n\geq0
\end{equation}
and the second limit relation in (\ref{b-cond}) holds with $m\geq m_\varepsilon$.

It follows from inequality (\ref{LAA-1}), the nonnegativity of $f_n$ and the assumption (\ref{a-b-assump}) that
$$
f_n(\rho_n)\geq \mu_n^m f_n([\Psi_{m}(\rho_n)])-a_f(1-\mu_n^m)\geq (1-\varepsilon)f_n([\Psi_{m}(\rho_n)])-\varepsilon\quad \forall n\geq0\;\, \forall  m\geq m_\varepsilon.
$$
So, the first limit relation in (\ref{b-cond}) implies that
\begin{equation}\label{u-b}
f_n(\rho_n)\leq C<+\infty\quad \textrm{and} \quad f_n([\Psi_{m}(\rho_n)])\leq 2C+1 \quad \forall n\geq0\;\, \forall  m\geq m_\varepsilon.
\end{equation}
It follows from the above inequalities that
\begin{equation}\label{e-way}
f_n([\Psi_{m}(\rho_n)])-f_n(\rho_n)\leq 2\varepsilon(C+1)\quad \forall n\geq0\;\,\forall m\geq m_\varepsilon.
\end{equation}
This implies  the first inequality in (\ref{A+}).

Since
$$
\mu_n^{m+k}[\Psi_{m+k}(\rho_n)]=\mu_n^{m}[\Psi_{m}(\rho_n)]+(\mu_n^{m+k}-\mu_n^{m})[\sigma_n^m]
$$
for any  $m\geq m_{\varepsilon}$, $k>0$ and $n\geq0$, where $\sigma_n^m=\Psi_{m+k}(\rho_n)-\Psi_{m}(\rho_n)$ is a positive operator, inequality (\ref{LAA-1}) and the nonnegativity of $f_n$ along with assumption (\ref{a-b-assump}) imply
\begin{equation}\label{m-k}
\begin{array}{rl}
 f_n([\Psi_{m+k}(\rho_n)])\!&\geq \,\frac{\mu_n^{m}}{\mu_n^{m+k}}f_n([\Psi_{m}(\rho_n)])-a_f\!\left(1-\frac{\mu_n^{m}}{\mu_n^{m+k}}\right)\\\\&\geq \, \mu_n^{m}f_n([\Psi_{m}(\rho_n)])-a_f(1-\mu_n^{m})\geq (1-\varepsilon)f_n([\Psi_{m}(\rho_n)])-\varepsilon.
\end{array}
\end{equation}

Note that $[\Psi_{m_\varepsilon}(\rho_n)]$ tends to $[\Psi_{m_\varepsilon}(\rho_0)]$ as $n\to+\infty$ because $m_\varepsilon\in M_{\rho_0}$.
By the limit relations in (\ref{b-cond}) there is $n_\varepsilon$ such that
$$
f_n(\rho_n)-f_0(\rho_0)\leq \varepsilon+A \quad \textrm{and} \quad f_0([\Psi_{m_\varepsilon}(\rho_0)])-f_n([\Psi_{m_\varepsilon}(\rho_n)])\leq \varepsilon \quad \forall n\geq n_\varepsilon.
$$
So, it follows from (\ref{m-e}), (\ref{u-b}) and (\ref{m-k}) with $m=m_\varepsilon$ and $k\geq0$ that
$$
f_n(\rho_n)-f_n([\Psi_{m_\varepsilon+k}(\rho_n)])\leq f_n(\rho_n)-f_n([\Psi_{m_\varepsilon}(\rho_n)])+\varepsilon(2C+2)\leq A+\varepsilon(2C+5)\quad \forall n\geq n_\varepsilon.
$$
This and (\ref{p-w-c}) imply the second inequality in (\ref{A+}).

To prove relation (\ref{B+}) note that inequality (\ref{LAA-1}) implies
$$
\Tr\Delta_m(\rho_n)f_n([\Delta_m(\rho_n)])\leq f_n(\rho_n)-\mu_n^m f_n([\Psi_{m}(\rho_n)])+a_f(1-\mu_n^m)\qquad \forall n\geq 0.
$$
This  allows us to derive (\ref{B+}) from (\ref{A+}) by using Lemma \ref{DTL-L} and the relations in (\ref{u-b}).

Assume that $\{g_n\}$ is a sequence of functions on $\S_0$ such that $|g_n(\rho)|\leq f_n(\rho)$ for any $\rho$ in $\S_0$. If all the functions $g_n$ satisfy inequality (\ref{LAA-1}) with the same function $a_g$, we have
$$
g_n([\Psi_m(\rho_n)])-g_n(\rho_n)\leq (1-\mu_n^m)(f_n([\Psi_m(\rho_n)])+f_n([\Delta_m(\rho_n)]))+a_g(1-\mu_n^m)
$$
for any $n\geq0$ and all $m$ large enough. Thus, Lemma \ref{DTL-L} and the second relation in (\ref{u-b})  allows us to derive (\ref{C-1+}) from (\ref{B+}).
If all the functions $g_n$ satisfy inequality (\ref{LAA-2}) with the same function $b_g$ then the functions $-g_n$ satisfy inequality (\ref{LAA-1}) with the function $a_{-g}=b_{g}$. So, (\ref{C-2+}) follows from (\ref{C-1+}). The last assertion is a corollary of the previous ones. $\square$

In the following sections we consider different applications of Proposition 1.

\section{Dominated convergence theorems and beyond}

\subsection{Basic dominated convergence theorem}

In study of characteristics of quantum systems and channels we often encounter the following situation:
if one characteristic $f$ majorizes the other characteristic $g$ then continuity of $f$ on some set of states implies  continuity of $g$.
In other words, there are many important  functions $f$ and $g$ on $\S(\H)$ related by  the inequality $|g(\rho)|\leq f(\rho)$ valid for any state $\rho$
such that
\begin{equation}\label{DCT-I-imp}
\left\{\exists \lim_{n\to+\infty}f(\rho_n)=f(\rho_0)<+\infty\right\}\quad\Rightarrow\quad\left\{\exists \lim_{n\to+\infty}g(\rho_n)=g(\rho_0)<+\infty\right\}
\end{equation}
for a sequence $\{\rho_n\}$ in $\S(\H)$ converging to a state $\rho_0$. Below we present several examples:
\begin{itemize}
  \item $f(\rho)=S(\rho_A)+S(\rho_B)$, $\rho\in\S(\H_{AB})\,$ and $\,g(\rho)=S(\rho)$ ($S$ is the von Neumann entropy);
  \item $f(\rho)=S(\rho)$, $\rho\in\S(\H_A)\,$ and $\,g(\rho)=I_c(\Phi,\rho)$ -- the coherent information of a quantum channel $\Phi:A\to B$ (defined in \eqref{CI-def} below);
  \item $f(\rho)=2S(\rho)$, $\rho\in\S(\H_A)\,$ and $\,g(\rho)=I(\Phi,\rho)$ -- the mutual information of a quantum channel $\Phi:A\to B$ (defined in \eqref{MI-Ch-def} below);
  \item $f(\rho)=S(\rho)+\ln k$, $\rho\in\S(\H_A)\,$ and $\,g(\rho)=S(\Phi(\rho))$ -- the output entropy a quantum channel $\Phi:A\to B$ with Choi rank $\leq k$;
  \item $f(\rho)=S(\rho)$, $\rho\in\S(\H_A)\,$ and $\,g(\rho)=ER(\M,\rho)$ -- the entropy reduction of a measurement over system $A$ described by POVM $\M$ (see \cite{L-ER,AMP,ER}).
\end{itemize}

The validity of implication (\ref{DCT-I-imp}) for the above functions $f$  and $g$ was originally proved by different methods. But it turns out that
in all the cases this implication is a corollary of part A of the following
theorem proved by using Proposition 1 in Section 3. \pagebreak

\begin{theorem}\label{DCT-I} \emph{Let $\S_0$ be a convex subset of $\,\S(\H)$ that
contains all finite rank states and $\{\rho_n\}$ a sequence in $\S_0$ converging to a state $\rho_0\in\S_0$.}\smallskip

 A) \emph{Let $f$ be a nonnegative lower semicontinuous function on $\S_0$ satisfying inequality \eqref{LAA-1}
and $g$ a function on $\S_0$ satisfying inequalities \eqref{LAA-1} and  \eqref{LAA-2} such that $|g(\rho)|\leq f(\rho)$ for any $\rho$ in $\S_0$.
If the function $g$ is bounded on the set of pure states then implication \eqref{DCT-I-imp} is valid.}\smallskip

B) \emph{Let $\{f_n\}_{n\geq 0}$ be a sequence of nonnegative lower semicontinuous functions on $\S_0$ satisfying inequality \eqref{LAA-1} with the same function $a_f$ (not depending on $n$) such that
\begin{equation}\label{l-s++}
\lim_{n\to+\infty}f_n(\rho_n)=f_0(\rho_0)<+\infty.
\end{equation}
Let   $\{g_n\}_{n\geq 0}$ be a sequence of functions on $\S_0$ satisfying inequalities \eqref{LAA-1} and  \eqref{LAA-2} with the same functions $a_g$  and $b_g$ (not depending on $n$) such that $|g_n(\rho)|\leq f_n(\rho)$ for any $\rho$ in $\S_0$ and all $\,n\geq0$. If
\begin{equation}\label{b-cond-I}
\liminf_{n\to+\infty} f_n(\rho'_n)\geq f_0(\rho'_0)\quad \textrm{and} \quad \lim_{n\to+\infty} g_n(\rho'_n)=g_0(\rho'_0)
\end{equation}
for any sequence $\{\rho'_n\}$ in $\S_0$ converging to a state $\rho'_0$ such that $\,\sup_n\rank\rho'_n<+\infty$ and $\rho'_n\leq 2\rho_n$ for all $n\geq0$ then
\begin{equation*}%\label{DCT-I-imp+}
\lim_{n\to+\infty}g_n(\rho_n)=g_0(\rho_0)<+\infty.
\end{equation*}
The second condition in \eqref{b-cond-I} can be replaced by the following one:
\begin{equation}\label{gen-con-I}
 \max\{|A-g_0(\rho'_0)|,|B-g_0(\rho'_0))|\}\leq G(C-f_0(\rho'_0)),
\end{equation}
where
$$
A=\liminf_{n\to+\infty} g_n(\rho'_n),\quad B=\limsup_{n\to+\infty} g_n(\rho'_n),\quad  C=\limsup_{n\to+\infty} f_n(\rho'_n)
$$
and $G(x)$ is a nonnegative nondecreasing function on $\mathbb{R}_+$ tending to $G(0)=0$ as $x\to0^+$.}
\end{theorem}\smallskip

\textbf{Remark 2.} Condition (\ref{gen-con-I}) is well defined, since the limit relation (\ref{l-s++}), the assumption  $\rho'_n\leq 2\rho_n$, $n\geq0$, and the validity of inequality  \eqref{LAA-1}
for all the functions $f_n$  imply that $|g_0(\rho'_0)|\leq f_0(\rho'_0)<+\infty$  and that $C<+\infty$. The first condition in \eqref{b-cond-I} shows that $C\geq0$.\smallskip

\emph{Proof.} Let
\begin{equation}\label{Psi-m-1}
  \Psi_m(\rho)=P_m^{\rho}\rho
\end{equation}
for any state $\rho$ in $\S(\H)$, where $P_m^{\rho}$ is the spectral projector of
$\rho$ corresponding to its $m$ maximal eigenvalues (taken the multiplicity into account).\footnote{See the remark after formula (83)
in \cite{QC} how to avoid the ambiguity of the definition of $P_m^{\rho}$  associated with multiple eigenvalues.} If $m>\rank \rho$ then $P_m^{\rho}$ is the projector onto the support of $\rho$.

It is shown at the end of Section 4.2.1 in \cite{QC} that the family $\{\Psi_m\}$ of maps from $\,\A=\S(\H)$ to $\T_+(\H)$
defined by formula \eqref{Psi-m-1} satisfies all the conditions stated at the begin of Section 3 provided that
\begin{equation}\label{M-s}
M_{\rho}=M^{s}_{\rho}\doteq\left\{m\in \mathbb{N}\,|\,\{\lambda^{\rho}_{m+1}< \lambda^{\rho}_m\}\vee\{\lambda^{\rho}_{m}=0\}\right\},
\end{equation}
where $\{\lambda^{\rho}_i\}$ is the sequence of eigenvalues of $\rho$ taken in the non-increasing order.
Denote by $\rho_n^m$ the state proportional to the operator $\Psi_m(\rho_n)$ for all sufficiently large $m$.
Note that $\rho_n^m\leq 2\rho_n$ for all $n\geq0$ and all $m$ large enough by Lemma \ref{DTL-L} in Section 3.\smallskip

A) In this case the function $g$ is continuous
on any set of states with bounded rank by Lemma 3 in \cite[Section 4.2.2]{QC}. It follows that
\begin{equation*}%\label{l-s+}
\lim_{n\to+\infty}g(\rho^m_n)=g(\rho^m_0)<+\infty\qquad \forall m\in M^s_{\rho_0}.
\end{equation*}
So, implication \eqref{DCT-I-imp} holds by the last claim of Corollary 7 in \cite[Section 4.2.1]{QC}.\smallskip

B) It suffices to consider the case when condition \eqref{gen-con-I} is met, since this condition holds trivially
if the second condition in \eqref{b-cond-I} is valid.

Note that  the finiteness of $g_0(\rho_0)$ follows from the assumed finiteness of $f_0(\rho_0)$.
By Proposition 1 (with $A=0$) the limit relation (\ref{l-s++}) and the first condition in \eqref{b-cond-I} imply existence of  vanishing sequences $\{\alpha_m\}$ and $\{\beta_m\}$ such that
\begin{equation}\label{f-g-u-a++}
  |f_n(\rho_n)-f_n(\rho^m_n)|\leq \alpha_m\quad \textrm{and} \quad |g_n(\rho_n)-g_n(\rho^m_n)|\leq \beta_m, \quad \forall n\geq0\;\; \forall m.
\end{equation}

Let $\varepsilon>0$ be arbitrary. Then there is  $m\in M_{\rho_0}$ such that
$G(2\alpha_m)+2\beta_m<\varepsilon$ and $\rho_n^m\leq 2\rho_n$ for all $n\geq0$. Limit relation \eqref{l-s++} and the first inequality in \eqref{f-g-u-a++}
imply that
$$
\limsup_{n\to+\infty} f_n(\rho^m_n)-f_0(\rho^m_0)\leq 2\alpha_m.
$$
By condition \eqref{gen-con-I} there $n_\varepsilon$ such that
$$
|g_n(\rho_n)-g_0(\rho_0)|\leq |g_n(\rho^m_n)-g_0(\rho^m_0)|+2\beta_m\leq G(2\alpha_m)+2\beta_m+\varepsilon<2\varepsilon\quad\forall n>n_\varepsilon,
$$
where the first inequality follows from the second inequality in \eqref{f-g-u-a++}. This implies
that $g_n(\rho_n)$ tends to $g_0(\rho_0)$ as $n\to+\infty$. $\square$\smallskip

By using Theorem \ref{DCT-I}A it is easy to show that implication \eqref{DCT-I-imp} holds for all the pairs of functions $f$ and $g$
presented before this theorem.\smallskip

\textbf{Example 1.} Consider
the functions $f(\rho)=S(\rho)$ and $g(\rho)=I_c(\Phi,\rho)$ -- the coherent information
of a quantum channel $\Phi:A\to B$ at a state $\rho$ in $\S(\H_A)$ with finite entropy defined
by the expression
\begin{equation}\label{CI-def}
I_c(\Phi,\rho)=I(B\!:\!R)_{\Phi\otimes\id_R(\hat{\rho})}-S(\rho),
\end{equation}
where $\hat{\rho}$ is pure state in $\S(\H_A\otimes\H_R)$ such that $\Tr_R\hat{\rho}=\rho$ \cite{H-SCI,Wilde}.
It is easy to show that $|g(\rho)|\leq f(\rho)$ for any state $\rho$ in the convex set $\,\S_0$ of states in $\S(\H_A)$ with finite entropy and that the function $g$  satisfies inequalities  \eqref{LAA-1} and \eqref{LAA-2} on $\,\S_0$ with $\,a_g=b_g=h_2$ \cite[Section 4.3]{MCB}.
So, it follows directly from Theorem \ref{DCT-I}A that implication \eqref{DCT-I-imp} holds for the functions $f$ and $g$.

Theorem \ref{DCT-I}B allows us to show  that
\begin{equation*}%\label{H-I}
\!\left\{\exists \lim_{n\to+\infty}S(\rho_n)=S(\rho_0)<+\infty\right\}\;\;\Rightarrow\;\;\left\{\exists \lim_{n\to+\infty}I_c(\Phi_n,\rho_n)=I_c(\Phi_0,\rho_0)<+\infty\right\}
\end{equation*}
for any sequence $\{\rho_n\}$ in $\S(\H_A)$ converging to a state $\rho_0$ and any sequence $\{\Phi_n\}$ of channels
strongly converging to a channel $\Phi_0$.\footnote{It means that $\Phi_n(\rho)$ tends to $\Phi_0(\rho)$ for any state $\rho$ in $\S(\H_A)$ \cite{CSR,Wilde+}.} Indeed, consider the families $\{f_n\}$ and $\{g_n\}$ of functions on $\S_0$, where
$f_n(\rho)=S(\rho)$ and $g_n(\rho)=I_c(\Phi_n,\rho)$ for all $n\geq0$. The first condition in \eqref{b-cond-I} holds by the lower semicontinuity of the entropy. To
prove the validity of the second  condition in \eqref{b-cond-I} note that for any sequence of states $\{\rho'_n\}$ in $\S_0$ converging to a state $\rho'_0$ such that $\,\sup_n\rank\rho'_n\leq d<+\infty\,$ there exists a sequence $\{\hat{\rho}'_n\}$ of pure states in $\S(\H_A\otimes \H_{R'})$ converging to a state $\hat{\rho}'_0$
such that $\Tr_{R'}\hat{\rho}'_n=\rho'_n$ for all $n\geq0$, where $\H_{R'}$ is a $d$-dimensional Hilbert space. The strong convergence
of the sequence $\{\Phi_n\}$ to the channel $\Phi_0$ implies that (cf.\cite{Wilde+})
$$
\lim_{n\to+\infty}\Phi_n\otimes\id_{R'}(\hat{\rho}'_n)=\Phi_0\otimes\id_{R'}(\hat{\rho}'_0).
$$
Thus  $g_n(\rho'_n)$ tends to $g_0(\rho'_0)$ by Remark 1 in Section 2.
%The implication \eqref{H-I} is originally proved in \cite{CMI} by completely different method.

\smallskip

Theorem \ref{DCT-I}B with condition \eqref{gen-con-I} is used essentially in the proof of Proposition 2
in Section 5.1.

\subsection{Simon-type dominated convergence theorem and preserving continuity under convex mixtures}

Simon's dominated convergence theorem for the von Neumann entropy (presented in \cite[the Appendix]{Ruskai})
can be formulated  (in the strengthened form obtained in \cite{SSP})  as follows: \emph{let $\{\rho_n\}$ and $\{\tau_n\}$
be sequences of states converging, respectively, to states $\rho_0$ and $\tau_0$ such that $\,c\rho_n\leq \tau_n$ for all $\,n\geq 0$ and some $c>0$, then}
\begin{equation}\label{S-DCT-II-imp}
\left\{\exists\lim_{n\to+\infty}S(\tau_n)=S(\tau_0)<+\infty\right\}\;\;\Rightarrow \;\;
\left\{\exists\lim_{n\to+\infty}S(\rho_n)=S(\rho_0)<+\infty \right\}.
\end{equation}
In fact, this implication is a partial case of the general result called Simon-type dominated convergence theorem in \cite{QC} (Theorem 9 in Section 4.2.3).
By this theorem  implication (\ref{S-DCT-II-imp}) holds for wide class of functions including important characteristics of quantum systems
and channels (bipartite and multipartite quantum mutual information, mutual information of a quantum channel as a function of input state, etc.)
\smallskip

Below we present an extended version of Theorem 9 in \cite{QC} designed for analysis of sequences of functions.\smallskip

\begin{theorem}\label{DCT-II}  \emph{Let $\S_0$ be a convex subset of $\,\S(\H)$ that
contains all finite rank states. Let $\{f_n\}_{n\geq 0}$ be a sequence of nonnegative lower semicontinuous functions on $\S_0$ satisfying inequalities (\ref{LAA-1}) and (\ref{LAA-2}) with the same functions $a_f$ and $b_f$ (not depending on $n$) such that
\begin{equation}\label{n-l-s-c}
\liminf_{n\to+\infty} f_n(\sigma_n)\geq f_0(\sigma_0)
\end{equation}
for any sequence $\{\sigma_n\}\subset\S_0$ converging to a state $\sigma\in\S_0$.}\smallskip

A) \emph{If $\{\tau_n\}\subset\S_0$ is a sequence converging to a state $\tau_0\in\S_0$ such that
\begin{equation}\label{DCT-II-tau}
f_0(\tau_0)<+\infty, \quad A\doteq\limsup_{n\to+\infty} f_n(\tau_n)-f_0(\tau_0)<+\infty
\end{equation}
and for a given $c>0$ the inequality
\begin{equation}\label{gen-con-II}
\limsup_{n\to+\infty} f_n(\rho'_n)-f_0(\rho'_0)\leq G_{\!c}\!\left(\limsup_{n\to+\infty} f_n(\tau'_n)-f_0(\tau'_0)\right)
\end{equation}
holds for any sequences $\{\rho'_n\}$ and $\{\tau'_n\}$ converging to states $\rho'_0$ and $\tau'_0$
such that $\,c\rho'_n\leq 2\tau'_n\leq 4\tau_n$ for all $\,n\geq0$ and $\,\sup_n\rank\tau'_n<+\infty$, where
$G_{\!c}(x)$ is a nondecreasing continuous function  on $\mathbb{R}_+$ such that $G_{\!c}(0)=0$,
then
\begin{equation}\label{DCT-II-rho}
f_0(\rho_0)<+\infty\quad \textit{and} \quad\limsup_{n\to+\infty}f_n(\rho_n)-f_0(\rho_0)\leq c^{-1}A+G_{\!c}(A)
\end{equation}
for any sequence $\{\rho_n\}\subset\S_0$ converging to a state $\rho_0\in\S_0$ such that $c\rho_n\leq \tau_n$ for all $\,n\geq0$.}\smallskip

B) \emph{If $\{\tau_n\}\subset\S_0$ is a sequence converging to a state $\tau_0\in\S_0$ such that
\begin{equation*}%\label{DCT-II-tau+}
\lim_{n\to+\infty} f_n(\tau_n)=f_0(\tau_0)<+\infty
\end{equation*}
and condition (\ref{gen-con-II}) holds  for a given $c>0$  then
\begin{equation*}%\label{DCT-II-tau++}
\lim_{n\to+\infty}f_n(\rho_n)=f_0(\rho_0)<+\infty
\end{equation*}
for any sequence $\{\rho_n\}\subset\S_0$ converging to a state $\rho_0\in\S_0$ such that $c\rho_n\leq \tau_n$ for all $\,n\geq0$.}\smallskip
\end{theorem}

\textbf{Remark 3.} Condition (\ref{gen-con-II}) is well defined, since the conditions in (\ref{DCT-II-tau}) and the assumed properties of the sequences
$\{\rho'_n\}$ and $\{\tau'_n\}$ imply, by the validity of inequality  \eqref{LAA-1}
for all the functions $f_n$, that $f_0(\sigma'_0)\leq\limsup_{n\to+\infty} f_n(\sigma'_n)<+\infty$, $\sigma=\rho,\tau$. By the proof of Theorem \ref{DCT-II} presented below  it suffices to require that condition (\ref{gen-con-II}) holds
for the sequences  $\{\rho'_n=\rho_n^m\}$ and $\{\tau'_n=\tau_n^m\}$ introduced in this proof (for all sufficiently large $m$).\smallskip

\emph{Proof.}  A) For a given operator $\rho$ in $\T_+(\H)$ denote by
$\widehat{m}(\rho)$ the maximal
number in the set
$M^{s}_{\rho}$ defined in (\ref{M-s}) not exceeding $m$ ($\widehat{m}(\rho)$ is well defined for all $m$ not less than the multiplicity of the maximal eigenvalue of $\rho$). For a  nonzero operator
$\sigma$ in $\T_+(\H)$ denote by  $P_{\widehat{m}(\rho)}^{\sigma}$ the spectral projector of $\sigma$ corresponding
to its $\widehat{m}(\rho)$ maximal eigenvalues.\footnote{See the remark after formula (83)
in \cite{QC} how to avoid the ambiguity of the definition of $P_{\widehat{m}(\rho)}^{\sigma}$  associated with multiple eigenvalues.}
If $\widehat{m}(\rho)>\rank \sigma$ then $P_{\widehat{m}(\rho)}^{\sigma}$ is the projector onto the support of $\sigma$.

Let $c>0$. Let $\{\rho_n\}$ and $\{\tau_n\}$  be sequences of states in $\S_0$ converging to states $\rho_0$ and $\tau_0$
in $\S_0$ such that $c\rho_n\leq \tau_n$ for all $\,n\geq0$ and the conditions in (\ref{DCT-II-tau}) and (\ref{gen-con-II}) hold.

Assume that $\A=\{\tau_n\}\cup\{\tau_0\}$ and
\begin{equation}\label{Psi-m-2}
  \Psi_m(\tau_n)=c P_{\widehat{m}(\rho_0)}^{\rho_n}\rho_n+P_{\widehat{m}(\sigma_0)}^{\sigma_n}\sigma_n, \quad n\geq0,
\end{equation}
where $\sigma_n=\tau_n-c\rho_n$, $P_{\widehat{m}(\rho_0)}^{\rho_n}$ and $P_{\widehat{m}(\sigma_0)}^{\sigma_n}$
are the spectral projectors of $\rho_n$ and $\sigma_n$ defined according to the above rule (it is implied that $\,P_{\widehat{m}(\sigma_0)}^{\sigma_n}\sigma_n=0\,$ if $\,\sigma_n=0$).%\footnote{We may assume w.l.o.g. that $\sigma_0\neq0$, so the projector $P_{\widehat{m}(\sigma_0)}^{\sigma_n}$ is well defined.}

It is shown at the end of Section 4.2.1 in \cite{QC} that the family  $\{\Psi_m\}$ of maps from
$\A=\{\tau_n\}\cup\{\tau_0\}$ to $\T_+(\H)$ defined by formula \eqref{Psi-m-2}
possesses all the properties stated at the begin of Section 3 with $M_{\tau_0}=\mathbb{N}\cap [m_*,+\infty)$, where $m_*$ is some positive integer.

Let $\mu_n^m=\Tr P_{\widehat{m}(\rho_0)}^{\rho_n}\rho_n$ and $\nu_n^m=\Tr\Psi_m(\tau_n)$ for all $n\geq0$ and $m\geq m_*$. By Lemma \ref{DTL-L} we have
\begin{equation}\label{mu-nu}
\lim_{m\to+\infty}\inf_{n\geq0}\mu^m_n=\lim_{m\to+\infty}\inf_{n\geq0}\nu^m_n=1.
\end{equation}
Denote by $\rho_n^m$, $\tau_n^m$ and $\sigma^m_n$ the states $(\mu^m_n)^{-1}P_{\widehat{m}(\rho_0)}^{\rho_n}\rho_n$,
$(\nu^m_n)^{-1}\Psi_m(\tau_n)$ and the operator $P_{\widehat{m}(\sigma_0)}^{\sigma_n}\sigma_n$ for all $n\geq0$ and all $m$ large enough.

Since $\tau_n=c\rho_n+\sigma_n$ and $\nu_n^m\tau_n^m=c\mu_n^m\rho_n^m+\sigma_n^m$, we have
\begin{equation}\label{imp-dec}
(1-\nu_n^m)[\tau_n-\nu_n^m\tau_n^m]=c(1-\mu_n^m)[\rho_n-\mu_n^m\rho_n^m]+(\sigma_n-\sigma_n^m),
\end{equation}
where the notation (\ref{BN}) is used.

By Proposition 1 in Section 3 relation \eqref{DCT-II-tau} and condition \eqref{n-l-s-c} imply existence of vanishing sequences $\{\alpha_m\}$ and $\{\beta_m\}$ such that
\begin{equation}\label{two-rel}
f_n(\tau_n^m)-f_n(\tau_n)\leq\alpha_m\quad \textrm{and} \quad (1-\nu_n^m)f_n([\tau_n-\nu_n^m\tau_n^m])\leq A+\beta_m\quad\forall n\neq 0\,\; \forall m.
\end{equation}
Due to relations (\ref{p-w-c}) and (\ref{p-w-c+}) applied to the function $f_0$ and the state $\tau_0$ we may assume that the above sequences are chosen in such a way that
\begin{equation}\label{two-rel-0}
|f_0(\tau_0)-f_0(\tau_0^m)|\leq \alpha_m\quad \textrm{and} \quad (1-\nu_0^m)|f_0([\tau_0-\nu_0^m\tau_0^m])|\leq \beta_m\quad \forall m.
\end{equation}

By relation \eqref{mu-nu} there is $m_0\geq m_*$ such that $\mu_n^m,\nu_n^m>1/2$ and hence $\rho_n^m\leq 2\rho_n$ and $\tau_n^m\leq 2\tau_n$ for all $n\geq0$ and all $m\geq m_0$. Assume in what follows that $m\geq m_0$ and $n\geq0$ are arbitrary. Since $\tau_n=c\rho_n+\sigma_n=c\mu_n^m\rho^m_n+c(\rho_n-\mu_n^m\rho^m_n)+\sigma_n$, by using the boundedness of the sequence $\{f_n(\tau_n)\}_{n\geq0}$ and inequality (\ref{LAA-1}) it is easy to show the boundedness of the sequence $\{f_n(\rho_n)\}_{n\geq0}$ and of the double sequence $\{f_n(\rho_n^m)\}_{n\geq0,\shs m\geq m_0}$. It implies, in particular, that $f_0(\rho_0)<+\infty$.  By using inequalities (\ref{LAA-1}) and (\ref{LAA-2}) it is easy to obtain
\begin{equation}\label{rho-n}
|f_n(\rho_n)-f_n(\rho_n^m)|\leq (1-\mu_n^m)(f_n([\rho_n-\mu_n^m\rho_n^m])+f_n(\rho_n^m))+(a_f+b_f)(1-\mu_n^m).
\end{equation}
Decomposition \eqref{imp-dec}, inequality (\ref{LAA-1}) and the nonnegativity
of $f_n$ imply that
\begin{equation}\label{tau-n}
c(1-\mu_n^m)f_n([\rho_n-\mu_n^m\rho_n^m])\leq(1-\nu_n^m)f_n([\tau_n-\nu_n^m\tau_n^m])+(1-\nu_n^m)a^*_f,
\end{equation}
where $a^*_f=\max_{x\in[0,1]}a_f(x)$. Note that (\ref{tau-n}) holds trivially if $\nu_n^m=1$, since in this case $\mu_n^m=1$.
The second inequalities in \eqref{two-rel} and  \eqref{two-rel-0} along with relation \eqref{mu-nu} and inequalities \eqref{rho-n} and \eqref{tau-n} imply, by the boundedness of the sequence $\{f_n(\rho_n^m)\}_{n\geq0,\shs m\geq m_0}$
mentioned before, the existence of a vanishing sequence $\{\gamma_m\}$ such that
\begin{equation}\label{gamma}
|f_0(\rho_0)-f_0(\rho_0^m)|\leq \gamma_m,\quad |f_n(\rho_n)-f_n(\rho_n^m)|\leq c^{-1}A+\gamma_m\quad \forall n\neq 0\quad\forall m\geq m_0.
\end{equation}

Since $m\in M_{\tau_0}$, the sequences $\{\tau^m_n\}_n$ and $\{\rho^m_n\}_n$ tend, respectively, to the states
$\tau^m_0$ and $\rho^m_0$. So, it follows from  \eqref{DCT-II-tau} and the first inequalities in \eqref{two-rel} and  \eqref{two-rel-0} that
\begin{equation}\label{tau-gap}
A_m\doteq\limsup_{n\to+\infty}f_n(\tau^m_n)-f_0(\tau^m_0)\leq A+2\alpha_m
\end{equation}
($A_m\geq 0$ by condition \eqref{n-l-s-c}). By  relation \eqref{mu-nu} and due to the  choice of $m_0$ we have  $\,4\tau_n\geq2\tau_n^m\geq 2c(\mu_n^m/\nu_n^m)\rho_n^m\geq c\rho_n^m\,$ for all $n\geq0$ and all sufficiently large $m>m_0$. Hence, the condition \eqref{gen-con-II} implies that
\begin{equation*}%\label{gen-con}
\limsup_{n\to+\infty}f_n(\rho^m_n)-f_0(\rho^m_0)\leq G_{\!c}(A_m)
\end{equation*}
for all such $m$. Thus, by using \eqref{gamma} and \eqref{tau-gap} we obtain
$$
\limsup_{n\to+\infty}f_n(\rho_n)-f_0(\rho_0)\leq c^{-1}A+2\gamma_m+G_{\!c}(A_m)\leq c^{-1}A+2\gamma_m+G_{\!c}(A+2\alpha_m)
$$
for all sufficiently large $m\geq m_0$. So, since $\{\alpha_m\}$ and $\{\gamma_m\}$ are vanishing sequences, the above inequality and the properties of the function $G_{\!c}$ imply the validity of the second relation in (\ref{DCT-II-rho}).\smallskip

B) This claim follows directly from claim  A (with $A=0$) due to condition (\ref{n-l-s-c}). $\square$\medskip

Theorem \ref{DCT-II}B  with $f_n=f$ coincides with  Theorem 9B in Section 4.2.3 in \cite{QC}, where it is applied to obtain dominated convergence
theorem for the quantum mutual information (Proposition 20A and Corollary 10 in \cite[Section 5.2.3]{QC}). Theorem \ref{DCT-II}B  is used essentially in the proofs of Proposition 2
in Section 5.1 and Proposition 4 in Section 5.2.
\smallskip

In what follows we will say that a double sequence $\{P^n_m\}_{n\geq0,m\geq m_0}$ ($m_0\in\mathbb{N}$) of finite rank projectors is \emph{consistent}
with a sequence $\{\rho_n\}\subset\T_+(\H)\setminus\{0\}$ converging to a nonzero operator $\rho_0$ if
\begin{equation}\label{P-prop}
\!\rank P^n_{m}\leq m,\;\; \Tr P^n_{m}\rho_n>0,\;\; P^n_{m}\leq P^n_{m+1}, \;\bigvee_{m\geq m_0}P^n_m\geq Q_n\;\; \textrm{and} \;\; s\textrm{-}\!\!\lim_{n\to+\infty}P^n_m=P^0_m\!
\end{equation}
for all $m\geq m_0$ and $n\geq0$, where  $Q_n$ is the projector onto the support of $\rho_n$ and the limit in the strong operator topology.\smallskip

The simplest example of a  double sequence of finite rank projectors consistent
with a given converging sequence $\{\rho_n\}$ is the sequence  $\{P^n_m\}_{n\geq0,m\geq m_0}$  such that $P^n_m=P_m$ for all $n\geq0$, where
$\{P_m\}_{m\geq m_0}$ is any increasing sequence of finite rank projectors strongly converging to the unit operator $I_{\H}$
such that $\Tr P_m\rho_n>0$ for all $m\geq m_0$ and $n\geq0$.\smallskip

Theorem \ref{DCT-II}A allows us to obtain  the following sufficient condition for convergence of $f_n(\rho_n)$ to $f_0(\rho_0)$
for a given sequence $\{f_n\}_{n\geq 0}$ of nonnegative lower semicontinuous LAA functions and
a sequence $\{\rho_n\}$ of states converging to a state $\rho_0$.
\smallskip

\begin{corollary}\label{DCT-II-c} \emph{Let $\S_0$ be a convex subset of $\,\S(\H)$ that
contains all finite rank states. Let $\{f_n\}_{n\geq 0}$ be a sequence of nonnegative lower semicontinuous functions on $\S_0$ satisfying inequalities (\ref{LAA-1}) and (\ref{LAA-2}) with the same functions $a_f$ and $b_f$ (not depending on $n$) such that condition (\ref{n-l-s-c}) holds
for any sequence $\{\sigma_n\}\subset\S_0$ converging to a state $\sigma\in\S_0$
and condition (\ref{gen-con-II}) with $c=1/2$ holds for any sequence $\{\tau_n\}\subset\S_0$ converging to a state $\tau_0\in\S_0$
and satisfying (\ref{DCT-II-tau}).}\smallskip

\emph{For each $n$ denote by $\tilde{f}_n$ the homogeneous
extension of $f_n$ to the cone generated by  $\S_0$ defined as}
\begin{equation}\label{h-ext}
\tilde{f}_n(\rho)=c_\rho f_n(\rho/c_\rho)\quad \textit{if}\quad c_\rho\doteq\Tr\rho\neq0 \quad\textit{and} \quad  \tilde{f}_n(0)=0.
\end{equation}

\emph{Let $\{\rho_n\}\subset\S_0$ be a sequence  converging to a state $\rho_0\in\S_0$.
If there is a double sequence $\{P^n_m\}_{n\geq0,m\geq m_0}$ of finite rank projectors
consistent  with the sequence $\{\rho_n\}$ (i.e. satisfying the conditions in (\ref{P-prop}))
such that
\begin{equation}\label{P-cont}
\lim_{n\to+\infty} \tilde{f}_n(P^n_m\rho_nP^n_m)=\tilde{f}_0(P^0_m\rho_0P^0_m)<+\infty\quad \forall m\geq m_0,
\end{equation}
all the operators $\bar{P}^n_m\rho_n\bar{P}^n_m$ belong to the cone generated by $\S_0$ and
\begin{equation}\label{P-cont+}
\lim_{m\to+\infty}\sup_{n\geq n_0} \tilde{f}_n(\bar{P}^n_m\rho_n\bar{P}^n_m)=0, \quad \textit{where} \;\;\bar{P}^n_m=I_{\H}-P^n_m,
\end{equation}
for some $n_0>0$ then}
\begin{equation}\label{P-cont++}
\lim_{n\to+\infty} f_n(\rho_n)=f_0(\rho_0)<+\infty.
\end{equation}
\end{corollary}

\emph{Proof.} Note first that the conditions in (\ref{P-prop}) imply that
\begin{equation*}%\label{pr-c}
  \lim_{n\to+\infty} P^n_m\rho_nP^n_m=P^0_m\rho_0P^0_m,\quad  \lim_{n\to+\infty} \bar{P}^n_m\rho_n\bar{P}^n_m=\bar{P}^0_m\rho_0\bar{P}^0_m\quad \textrm{and}   \;\;\lim_{m\to+\infty}\Tr\bar{P}^n_m\rho_n=0
\end{equation*}
for all $m$ and $n$.
So, by condition (\ref{n-l-s-c}) the validity of (\ref{P-cont+}) implies that
\begin{equation}\label{P-cont+0}
\lim_{m\to+\infty}\tilde{f}_0(\bar{P}^0_m\rho_0\bar{P}^0_m)=0, \quad \textup{where} \;\;\bar{P}^0_m=I_{\H}-P^0_m.
\end{equation}
It follows from inequalities (\ref{LAA-1}) and (\ref{LAA-2}) that for any given $m\geq m_0$ we have
\begin{equation}\label{0-in+}
 f_0(P^0_m\rho_0P^0_m+\bar{P}^0_m\rho_0\bar{P}^0_m)\geq \tilde{f}_0(P^0_m\rho_0P^0_m)+\tilde{f}_0(\bar{P}^0_m\rho_0\bar{P}^0_m)-a_f(\Tr\bar{P}^0_m\rho_0)
\end{equation}
and
\begin{equation}\label{n-in+}
 f_n(P^n_m\rho_nP^n_m+\bar{P}^n_m\rho_n\bar{P}^n_m)\leq \tilde{f}_n(P^n_m\rho_nP^n_m)+\tilde{f}_n(\bar{P}^n_m\rho_n\bar{P}^n_m)+b_f(\Tr\bar{P}^n_m\rho_n)
\end{equation}
for all $n\geq0$. Inequality (\ref{n-in+}) with $n=0$ shows, due to relations  (\ref{P-cont}) and (\ref{P-cont+0}), the finiteness of   $f_0(P^0_m\rho_0P^0_m+\bar{P}^0_m\rho_0\bar{P}^0_m)$.
Since $\Tr\bar{P}^n_m\rho_n$ tends to zero as $m\to+\infty$ uniformly on $n\geq0$ by Dini's lemma, inequalities (\ref{0-in+}) and  (\ref{n-in+}) along with relations (\ref{P-cont}),(\ref{P-cont+}) and (\ref{P-cont+0})
imply that
\begin{equation}\label{n-in}
 \limsup_{n\to+\infty} f_n(P^n_m\rho_nP^n_m+\bar{P}_m\rho_n\bar{P}_m)\leq f_0(P^0_m\rho_0P^0_m+\bar{P}^0_m\rho_0\bar{P}^0_m)+\varepsilon_m,\quad \forall m,
\end{equation}
where  $\{\varepsilon_m\}$  is a sequence tending to zero as $\,m\to+\infty$.\smallskip

Since $P^n_m\rho_nP^n_m+\bar{P}^n_m\rho_n\bar{P}^n_m=\frac{1}{2}(\rho_n+[U^n_m]^*\rho_n U^n_m)$, where $U^n_m=2P^n_m-I_{\H}$ is a unitary operator,
we have $P^n_m\rho_nP^n_m+\bar{P}^n_m\rho_n\bar{P}^n_m\geq\frac{1}{2}\rho_n$ for all $n\geq0$. Thus, by applying Theorem \ref{DCT-II}A and using (\ref{n-in})
we obtain
\begin{equation}\label{0-in}
 \limsup_{n\to+\infty} f_n(\rho_n)-f_0(\rho_0)\leq 2\varepsilon_m+G_{1/2}(\varepsilon_m) \quad \forall m,
\end{equation}
where $G_{1/2}$ is the function that exists due to the assumption that condition (\ref{gen-con-II}) holds with $c=1/2$.
Since the r.h.s. of (\ref{0-in}) tends to zero as $m\to+\infty$, this relation implies (\ref{P-cont++}) due to the assumed validity of condition (\ref{n-l-s-c}) for the sequence $\{f_n\}$. $\Box$
\medskip

The following theorem is a generalization of Theorem 10 in \cite[Section 4.2.3]{QC} for sequences of LAA functions.\smallskip

\begin{theorem}\label{convex-m}
\emph{Let $\S_0$ be a convex subset of $\,\S(\H)$ and $\widetilde{\S}_0$ the cone generated by $\S_0$. Let $\{f_n\}_{n\geq 0}$ be a sequence of nonnegative lower semicontinuous functions on $\S_0$ satisfying inequalities (\ref{LAA-1}) and (\ref{LAA-2}) with the same functions $a_f$ and $b_f$ (not depending on $n$) such that
\begin{equation}\label{f-cond}
\liminf_{n\to+\infty}f_n(\varrho_n)\geq f_0(\varrho_0)
\end{equation}
for any sequence $\{\varrho_n\}\subset\S_0$  converging to a state  $\varrho_0\in\S_0$. Let $\,\{\rho_n\}$ and $\,\{\sigma_n\}$ be sequences of states in $\S_0$ converging, respectively,
to states  $\rho_0$ and $\sigma_0$ in $\S_0$ such that
\begin{equation}\label{23-two}
\lim_{n\to+\infty} f_n(\rho_n)=f_0(\rho_0)<+\infty\quad\textrm{and}\quad\lim_{n\to+\infty} f_n(\sigma_n)=f_0(\sigma_0)<+\infty.
\end{equation}
If there exist  families $\{\Psi_m^{\rho}\}$ and $\{\Psi_m^{\sigma}\}$ of maps from $\,\A_{\rho}\doteq\{\rho_n\}_{n\geq0}$ and $\,\A_{\sigma}\doteq\{\sigma_n\}_{n\geq0}$ to $\widetilde{\S}_0$
satisfying the conditions stated at the begin of Section 3 with the sets $M_{\rho_0}$ and $M_{\sigma_0}$  such that $\,\mathrm {card}(M_{\rho_0}\cap M_{\sigma_0})=+\infty\,$ and
\begin{equation}\label{sum-rel}
\lim_{n\to+\infty}f_n(p_n [\Psi_m^{\rho}(\rho_n)]+\bar{p}_n[\Psi_m^{\sigma}(\sigma_n)])=f_0(p_0 [\Psi_m^{\rho}(\rho_0)]+\bar{p}_0[\Psi_m^{\sigma}(\sigma_0)])<+\infty
\end{equation}
for any sequence $\,\{p_n\}$ in $[0,1]$  converging to $p_0$ and all sufficiently large $m$ in $M_{\rho_0}\cap M_{\sigma_0}$, where $\,\bar{p}_n=1-p_n$ and $\,[\varrho]$ denotes the state proportional to a positive operator $\varrho$, then
\begin{equation}\label{23-one}
\lim_{n\to+\infty}f_n(p_n\rho_n+\bar{p}_n\sigma_n)=f_0(p_0\rho_0+\bar{p}_0\sigma_0)<+\infty,\quad  \bar{p}_n=1-p_n,
\end{equation}
for any sequence $\,\{p_n\}$  in $[0,1]$  converging to $p_0$.
}\end{theorem}\smallskip\smallskip

\emph{Proof.} We will follow the proof of Theorem 10 in \cite{QC} with necessary modifications.\smallskip

Let  $\mu_n^m=\Tr \Psi_m^{\rho}(\rho_n)$ and $\nu_n^m=\Tr \Psi_m^{\sigma}(\sigma_n)$.  By Lemma \ref{DTL-L} in Section 3 we have
\begin{equation}\label{mu-nu+gc}
\lim_{m\to+\infty}\inf_{n\geq0}\mu^m_n=\lim_{m\to+\infty}\inf_{n\geq0}\nu^m_n=1.
\end{equation}
Denote by $\rho_n^m$ and $\sigma_n^m$ the states $(\mu_n^m)^{-1}\Psi_m^{\rho}(\rho_n)$ and $(\nu_n^m)^{-1}\Psi_m^{\sigma}(\sigma_n)$ for all sufficiently large $m$.

By applying Proposition 1 in Section 3 (with $A=0$)
to the sequence $\{f_n\}$ (which is possible due to condition (\ref{f-cond})) we conclude that the limit relations in \eqref{23-two}
imply that
\begin{equation}\label{2-r-e-gc}
 \lim_{m\to+\infty}\sup_{n\geq0}(1-\mu_n^m)f_n([\rho_n-\mu_n^m\rho_n^m])=\lim_{m\to+\infty}\sup_{n\geq0}(1-\nu_n^m)f_n([\sigma_n-\nu_n^m\sigma_n^m])=0
\end{equation}
and that
\begin{equation}\label{2-rel+gc}
\sup_{n\geq0}f_n(\rho^m_n),\; \sup_{n\geq0}f_n(\sigma^m_n)\leq C<+\infty
\end{equation}
for all sufficiently large $m$.

Let $\vartheta_n=p_n\rho_n+\bar{p}_n\sigma_n$ for a given arbitrary sequence $\,\{p_n\}\subset[0,1]$  converging to $p_0\in[0,1]$. Consider the family $\Psi^{\vartheta}_m$ of maps from $\,\A_{\vartheta}\doteq\{\vartheta_n\}_{n\geq0}$ to $\widetilde{\S}_0$ defined as
$$
\Psi^{\vartheta}_m(\vartheta_n)=p_n\Psi^{\rho}_m(\rho_n)+\bar{p}_n\Psi^{\sigma}_m(\sigma_n)=p_n\mu_n^m\rho^m_n+\bar{p}_n\nu_n^m\sigma^m_n
$$
It satisfies all the conditions stated at the begin of Section 3 with
$M_{\vartheta_0}\doteq M_{\rho_0}\cap M_{\sigma_0}$.

It follows from (\ref{mu-nu+gc}) that condition \eqref{sum-rel} implies
\begin{equation}\label{2-rel++gc}
\lim_{n\to+\infty}f_n([\Psi^{\vartheta}_m(\vartheta_n)])=f_0([\Psi^{\vartheta}_m(\vartheta_0)])<+\infty
\end{equation}
for all sufficiently large $m$ in $M_{\vartheta_0}$.

Since all the functions $f_n$ satisfy  inequality \eqref{LAA-1} with the function $a_f$ and inequality \eqref{LAA-2} with the function $b_f$, we have
$$
\begin{array}{ccc}
  |f_n(\vartheta_n)-f_n([\Psi^{\vartheta}_m(\vartheta_n)])|\leq \delta^m_n (f_n([\vartheta_n-\Psi^{\vartheta}_m(\vartheta_n)])+f_n([\Psi^{\vartheta}_m(\vartheta_n)]))+(a_f+b_f)(\delta^m_n)\\\\
  \leq p_n(1-\mu_n^m)f_n([\rho_n-\mu_n^m\rho_n^m])+\bar{p}_n(1-\nu_n^m)f_n([\sigma_n-\nu_n^m\sigma_n^m])+\delta^m_nb^*_f\\\\
  +p_n\mu_n^m \delta^m_n(1-\delta^m_n)^{-1}f_n(\rho_n^m)+\bar{p}_n\nu_n^m\delta^m_n(1-\delta^m_n)^{-1}f_n(\sigma_n^m)+\delta^m_nb^*_f+(a_f+b_f)(\delta^m_n),
\end{array}
$$
where $\delta^m_n=1-p_n\mu^m_n-\bar{p}_n\nu^m_n$ and $b^*_f=\max_{x\in[0,1]}b_f(x)$, for all $n\geq0$. Thus, it follows from \eqref{mu-nu+gc}, \eqref{2-r-e-gc}  and \eqref{2-rel+gc} that
$$
 \lim_{m\to+\infty,\, m\in M_{\vartheta_0}}\sup_{n\geq0}|f_n(\vartheta_n)-f_n([\Psi^{\vartheta}_m(\vartheta_n)])|=0.
$$
This and \eqref{2-rel++gc} imply \eqref{23-one}. $\square$\smallskip

Theorem \ref{convex-m}  is used in the proofs of Proposition 3
in Section 5.1 and Proposition 4 in Section 5.2.

\section{Applications to characteristics of quantum systems and channels}

\subsection{The quantum relative entropy}

\subsubsection{Lindblad's extension of the quantum relative entropy and its approximation}
The quantum relative entropy for two states $\rho$ and
$\sigma$ in $\mathfrak{S}(\mathcal{H})$ defined in \eqref{qre-def} can be extended to any
operators $\rho$ and
$\sigma$ in $\mathfrak{T}_+(\mathcal{H})$ in different ways. We will use the Lindblad's extension
defined as
\begin{equation}\label{qre-def+}
D(\rho\shs\|\shs\sigma)=\sum_i\langle
i|\,\rho\ln\rho-\rho\ln\sigma\,|i\rangle+\Tr\sigma-\Tr\rho,
\end{equation}
where $\{|i\rangle\}$ is the orthonormal basis of
eigenvectors of the operator  $\rho$ and it is assumed that $\,D(0\|\shs\sigma)=\Tr\sigma\,$ and
$\,D(\rho\shs\|\sigma)=+\infty\,$ if $\,\mathrm{supp}\rho\shs$ is not
contained in $\shs\mathrm{supp}\shs\sigma$ (in particular, if $\rho\neq0$ and $\sigma=0$)
\cite{L-2}.\smallskip

The function $(\rho,\sigma)\mapsto D(\rho\shs\|\shs\sigma)$ is nonnegative lower semicontinuous and jointly convex on
$\T_+(\H)\times\T_+(\H)$. It has the following properties:
\begin{itemize}
  \item for any $\rho,\sigma\in\T_+(\H)$ and $c\geq0$ the following equalities  hold:
  \begin{equation}\label{D-mul}
  D(c\rho\shs\|\shs c\sigma)=cD(\rho\shs\|\shs \sigma),\qquad\qquad\qquad\qquad\quad\;\;
  \end{equation}
\begin{equation}\label{D-c-id}
D(\rho\shs\|\shs c\sigma)=D(\rho\shs\|\shs\sigma)-\Tr\rho\ln c+(c-1)\Tr\sigma;
\end{equation}
  \item for any operators $\rho$, $\sigma$ and $\omega$ in $\T_+(\H)$ the following inequalities hold (with possible values $+\infty$ in one or both sides)
  \begin{equation}\label{re-2-ineq-conc}
  D(\rho+\sigma\|\shs\omega)\geq D(\rho\shs\|\shs\omega)+D(\sigma\shs\|\shs\omega)-\Tr\omega,\qquad\qquad\qquad\quad\;\,
  \end{equation}
  \begin{equation}\label{re-2-ineq-conv}
  D(\rho+\sigma\|\shs\omega)\leq D(\rho\shs\|\shs\omega)+D(\sigma\shs\|\shs\omega)+ H(\{\Tr\rho,\Tr\sigma\})-\Tr\omega,
  \end{equation}
  where $H(\{\Tr\rho,\Tr\sigma\})$ is the extended binary  entropy of $\{\Tr\rho,\Tr\sigma\}$ defined after (\ref{w-k-ineq+}).\footnote{If the extended von Neumann entropy of the operators $\rho$ and $\sigma$ is finite then inequalities (\ref{re-2-ineq-conc}) and (\ref{re-2-ineq-conv}) follow, due to representation (\ref{re-exp}), from the inequalities in (\ref{w-k-ineq+}). In the general case these inequalities can be proved by approximation using Lemma 4 in \cite{L-2}.}
 \end{itemize}

If the extended von Neumann entropy $S(\rho)$ of $\rho$ (defined in (\ref{S-ext})) is finite
then
\begin{equation}\label{re-exp}
D(\rho\shs\|\shs\sigma)=\Tr\rho(-\ln\sigma)-S(\rho)-\eta(\Tr\rho)+\Tr\sigma-\Tr\rho,
\end{equation}
where $\rho\mapsto\Tr\rho(-\ln\sigma)$ is an affine lower semicontinuous nonnegative function on $\T_+(\H)$ defined according to the rule (\ref{H-as}).

Representation \eqref{re-exp} is quite useful. It allows, in particular, to prove continuity of the function $\rho\mapsto D(\rho\shs\|\shs\sigma)$ on any subset of $\T_+(\H)$
where the function $\rho\mapsto\Tr\rho(-\ln\sigma)$ is continuous. This  follows from the lower semicontinuity of the entropy and the relative entropy.

It follows from Proposition 5.24 in \cite{O&P} that for any nonzero operator $\sigma$ in $\T_+(\H)$ the function $f(\rho)=D(\rho\shs\|\shs\sigma)$ is a lower semicontinuous nonnegative convex function on $\S(\H)$  satisfying inequality (\ref{LAA-1}) with $a_f=h_2$. So, if $\{\Psi_m\}$
is a family of maps from a given subset $\A\subseteq \S(\H)$ into $\T_+(\H)$ with the properties stated at the begin of Section 3 then Proposition 1 in \cite[Section 4.2.1]{QC} shows that
\begin{equation}\label{re-ap}
  D(\rho\shs\|\shs\sigma)=\lim_{m\to+\infty}D([\Psi_m(\rho)]\|\shs\sigma),
\end{equation}
where $[\Psi_m(\rho)]=(\Tr\Psi_m(\rho))^{-1}\Psi_m(\rho)$, for any state $\rho$ in $\A$ and that this convergence is uniform on any compact subset of $\A$ on which the function $\rho\mapsto D(\rho\shs\|\shs\sigma)$ is continuous.

If $\A=\S(\H)$ and $\{\Psi_m\}$ is a family defined in (\ref{Psi-m-1}) then $\,S([\Psi_m(\rho)])\leq \ln m\,$ for any state $\rho$ and hence one can rewrite \eqref{re-ap} as follows
\begin{equation}\label{re-ap+}
  D(\rho\shs\|\shs\sigma)=\lim_{m\to+\infty}(\Tr[\Psi_m(\rho)](-\ln\sigma)-S([\Psi_m(\rho)]))+\Tr\sigma-1\quad  \forall\rho\in\S(\H).
\end{equation}

As a simple example of using  approximation relation \eqref{re-ap+} consider the proof of the inequality
\begin{equation}\label{re-ineq}
  D(\rho\shs\|\shs\sigma+\vartheta)\leq D(\rho\shs\|\shs\sigma)+\Tr\vartheta
\end{equation}
valid for any operators $\rho$, $\sigma$ and $\vartheta$ in $\T_+(\H)$. Identity (\ref{D-mul}) allows us to assume that
$\rho$ is a state (if $\rho=0$ then (\ref{re-ineq}) holds trivially). If $S(\rho)$
is finite then the validity of \eqref{re-ineq} directly follows from representation \eqref{re-exp}, since
$$
-\Tr\rho\ln(\sigma+\vartheta)\leq-\Tr\rho\ln \sigma
$$
by the operator monotonicity of the logarithm. As $S([\Psi_m(\rho)])\leq \ln m$, relation \eqref{re-ap+} allows us to prove the validity of \eqref{re-ineq} for arbitrary state $\rho$ (with possible values $+\infty$ in both sides).

Assume now that $\{(\rho_n,\sigma_n)\}$ is a sequences of pairs in $\S(\H)\times\T_+(\H)$ converging, respectively,
to a pair $(\rho_0,\sigma_0)$ such that
\begin{equation}\label{D-conv}
\lim_{n\to+\infty}D(\rho_n\|\shs\sigma_n)=D(\rho_0\|\shs\sigma_0)<+\infty.
\end{equation}
Then Proposition 1 in Section 3 with the sequence of functions $f_n(\varrho)=D(\varrho\shs\|\shs\sigma_n)$ shows that
\begin{equation*}%\label{n-u-a}
\lim_{m\to+\infty} \sup_{n\geq0}\left |D(\rho_n\shs\|\shs\sigma_n)-D([\Psi_m(\rho_n)]\|\shs\sigma_n)\right|=0,
\end{equation*}
i.e. the values of the relative entropy at the sequences
$\{(\rho_n,\sigma_n)\}$ and $\{([\Psi_m(\rho_n)],\sigma_n)\}$
are uniformly close to each other for large $m$. This gives a way to study conditions under which relation (\ref{D-conv}) holds.
Below we consider several results in this direction  obtained by the above approximation technique.

\subsubsection{Dominated-type continuity conditions.}  The following proposition gives the relative entropy version of the
Simon's dominated convergence theorem for the von Neumann  entropy mentioned as the begin of Section 4.2. \smallskip

\textbf{Proposition 2.}  \emph{Let $\,\{\rho^1_n\}$, $\{\rho^2_n\}$, $\{\sigma^1_n\}$ and $\,\{\sigma^2_n\}$ be sequences of operators in $\T_+(\H)$ converging, respectively,
to operators  $\rho^1_0$, $\rho^2_0$, $\sigma^1_0$ and $\sigma^2_0$ such that $\,\rho^2_n\leq \rho^1_n$ and $\,\sigma^1_n\leq \sigma^2_n$ for all $\,n\geq0$. If there exists
\begin{equation}\label{c-11+}
\lim_{n\to+\infty}D(\rho^1_n\|\shs\sigma^1_n)=D(\rho^1_0\|\shs\sigma^1_0)<+\infty
\end{equation}
then}
\begin{equation}\label{c-22+}
\lim_{n\to+\infty}D(\rho^2_n\|\shs\sigma^2_n)=D(\rho^2_0\|\shs\sigma^2_0)<+\infty.
\end{equation}

\textbf{Remark 4.} The conditions $\,\rho^2_n\leq \rho^1_n$ and $\,\sigma^1_n\leq \sigma^2_n$ for all $\,n\geq0$
in Proposition 2 can be replaced by the conditions $\,c_{\rho}\rho^2_n\leq \rho^1_n$ and $\,c_{\sigma}\sigma^1_n\leq \sigma^2_n$ for all $\,n\geq0$ and some $c_{\rho},c_{\sigma}>0$. This can be shown by using identities (\ref{D-mul}) and (\ref{D-c-id}).\smallskip

\emph{Proof.} At the first step we will prove that (\ref{c-11+}) implies (\ref{c-22+}) provided that
$\sigma^1_0\neq0$ and that $\,\{\rho^1_n\}$ and $\{\rho^2_n\}$ are sequences of states in $\S(\H)$ converging, respectively,
to states $\rho^1_0$ and $\rho^2_0$ such that $\,c\rho^2_n\leq \rho^1_n$ for all $\,n\geq0$ and some $c>0$. In this case we may assume
that all the operators $\sigma^1_n$ and $\sigma^2_n$ are nonzero for all $n\geq0$.

Assume that (\ref{c-11+}) holds. Consider the sequences of functions $f_n(\rho)=D(\rho\shs\|\shs\sigma^1_n)+\Tr(\sigma^2_n-\sigma^1_n)$ and $g_n(\rho)=D(\rho\shs\|\shs\sigma^2_n)$ on $\S(\H)$, $n\geq0$. All these functions are
nonnegative lower semicontinuous convex and satisfy inequality \eqref{LAA-1} with the same function $a_f=h_2$ \cite[Proposition 5.24]{O&P}. By inequality \eqref{re-ineq}
we have $g_n(\rho)\leq f_n(\rho)$ for any state $\rho\in\S(\H)$ and all $n\geq0$.  So, we will prove first that \eqref{c-11+}
implies the existence of
\begin{equation}\label{c-12+}
 \lim_{n\to+\infty}D(\rho^1_n\|\shs\sigma^2_n)=D(\rho^1_0\|\shs\sigma^2_0)<+\infty
\end{equation}
by using Theorem \ref{DCT-I}B in Section 4.1. The first condition in \eqref{b-cond-I}
for the sequence $\{f_n\}$ follows from the lower semicontinuity of the function
\begin{equation}\label{j-l-s}
(\rho,\sigma)\mapsto D(\rho\shs\|\shs\sigma).
\end{equation}
To show the validity of condition \eqref{gen-con-I} with the function $G(x)=x\,$ it suffices, by the lower semicontinuity
of the function \eqref{j-l-s}, to show that
$$
\limsup_{n\to+\infty}D(\rho'_n\|\shs\sigma^2_n)-D(\rho'_0\|\shs\sigma^2_0)\leq \limsup_{n\to+\infty} D(\rho'_n\|\shs\sigma^1_n)-D(\rho'_0\|\shs\sigma^1_0)
$$
for any sequence $\{\rho'_n\}$ of states with bounded rank converging to a state $\rho'_0$ such that $\rho'_n\leq 2\rho_n$ for all $n\geq0$. Since
for any such sequence we  have
\begin{equation}\label{H-lim}
\lim_{n\to+\infty}S(\rho'_n)=S(\rho_0')<+\infty,
\end{equation}
this can be done by using representation \eqref{re-exp} and Proposition 5 in the Appendix.

To show that \eqref{c-12+} implies \eqref{c-22+} it suffices to  apply Theorem \ref{DCT-II}B in Section 4.2 to the sequence of functions $f_n(\rho)=D(\rho\shs\|\shs\sigma^2_n)$
with the sequences of states $\tau_n=\rho^1_n$ and $\rho_n=\rho^2_n$. Indeed, in this case the condition \eqref{n-l-s-c}
holds by the lower semicontinuity of the function \eqref{j-l-s}. Since for any sequence $\{\rho'_n\}$ of states with bounded rank converging to a state $\rho'_0$
relation \eqref{H-lim} holds, by using representation \eqref{re-exp} and Proposition 5 in the Appendix it is easy to show
that  condition \eqref{gen-con-II} holds with the function $G_{\!c}(x)=x/c$.

By using the result proved at the first step and identity (\ref{D-mul}) it is easy to establish the claim
of the proposition in the case when all the limit operators $\rho^1_0$, $\rho^2_0$,  $\sigma^1_0$ and $\sigma^2_0$ are nonzero.

To prove that (\ref{c-11+}) implies (\ref{c-12+}) if  $\rho^1_0=0$ and $\sigma^1_0$ and $\sigma^2_0$ are arbitrary note that
inequality (\ref{re-ineq}) gives
$$
D(\rho^1_n\|\shs\sigma^2_n)\leq D(\rho^1_n\|\shs\sigma^1_n)+\Tr(\sigma^2_n-\sigma^1_n)\quad \forall n
$$
and hence (\ref{c-11+}) shows that
$$
\limsup_{n\to+\infty}D(\rho^1_n\|\shs\sigma^2_n)\leq D(0\|\shs\sigma^1_0)+\Tr(\sigma^2_0-\sigma^1_0)=\Tr\sigma^2_0=D(0\|\shs\sigma^2_0).
$$
By the lower semicontinuity of the relative entropy, the last inequality  implies (\ref{c-12+}).

This completes the proof of the implication (\ref{c-11+})$\Rightarrow$(\ref{c-12+}), since the assumption $\rho^1_0\neq0$ and $\sigma_0^1=0$
contradicts to (\ref{c-11+}).

To prove that (\ref{c-12+}) implies (\ref{c-22+}) in the general case note that
\begin{equation}\label{D-12}
 D(\rho_n^2\|\shs\sigma^2_n)+D(\delta_n\|\shs\sigma^2_n)\leq D(\rho_n^1\|\shs\sigma^2_n)+\Tr\sigma^2_n,\quad \forall n\geq0,
\end{equation}
where $\delta_n=\rho_n^1-\rho_n^2$, by inequality (\ref{re-2-ineq-conc}).

If $\,\rho^1_0\neq0$, $\rho^2_0=0$, $\sigma^1_0\neq0\,$ then the sequence $\{\delta_n\}$ tends to $\rho^1_0$. So, as $\delta_n\leq\rho^1_n$ for all $n\,$ and $\,\rho^1_0\neq0$, relation (\ref{c-12+}) implies, by the above part of the proof, that
$$
\lim_{n\to+\infty}D(\delta_n\|\shs\sigma^2_n)=D(\rho^1_0\shs\|\shs\sigma^2_0)<+\infty.
$$
Thus, inequality (\ref{D-12}) and  relation (\ref{c-12+}) show that
\begin{equation}\label{D-21}
\limsup_{n\to+\infty}D(\rho_n^2\|\shs\sigma^2_n)\leq \Tr\sigma^2_0=D(0\shs\|\shs\sigma^2_0).
\end{equation}
By the lower semicontinuity of the relative entropy, the last inequality  implies (\ref{c-22+}).

If $\rho^1_0=\rho^2_0=0$ and $\sigma^1_0$ is arbitrary then inequality (\ref{D-12}) and  relation (\ref{c-12+}) show that
$$
\limsup_{n\to+\infty}D(\rho_n^2\|\shs\sigma^2_n)+\liminf_{n\to+\infty}D(\delta_n\|\shs\sigma^2_n)\leq\limsup_{n\to+\infty}(D(\rho_n^2\|\shs\sigma^2_n)
+D(\delta_n\|\shs\sigma^2_n))\leq 2\Tr\sigma^2_0=2D(0\shs\|\shs\sigma^2_0).
$$
Since the sequence $\{\delta_n\}$ tends to $\rho_0^1-\rho_0^2=0$, the last inequality and
the lower semicontinuity of the relative entropy imply relation (\ref{D-21}) and hence the validity of (\ref{c-22+}).
$\square$\smallskip

\begin{corollary}\label{re-br+++}  \emph{Let $\,\{\rho_n\}$ and  $\{\sigma_n\}$ be sequences of operators in $\T_+(\H)$ converging, respectively,
to operators  $\rho_0$ and $\sigma_0$. If there exist operators $\rho_*$ and $\sigma_*$ in $\T_+(\H)$ such that  $\,\rho_n\leq \rho_*$,  $\,\sigma_*\leq \sigma_n$ for all $\,n\geq0$ and $D(\rho_*\|\shs\sigma_*)<+\infty$ then}
\begin{equation*}%\label{c-11}
\lim_{n\to+\infty}D(\rho_n\|\shs\sigma_n)=D(\rho_0\|\shs\sigma_0)<+\infty.
\end{equation*}
\end{corollary}

\subsubsection{Preserving local continuity under summation.}

It is shown in \cite{SSP} that
\begin{equation*}%
\lim_{n\to+\infty}S(\rho_n+\sigma_n)=S(\rho_0+\sigma_0)<+\infty
\end{equation*}
for any  sequences $\{\rho_n\}$ and $\{\sigma_n\}$  of operators in $\,\T_+(\H)$ converging
to operators  $\rho_0$ and $\sigma_0$ provided that
\begin{equation*}%\label{2-rel}
\lim_{n\to+\infty}S(\rho_n)=S(\rho_0)<+\infty\quad \textup{and}
\quad\lim_{n\to+\infty}S(\sigma_n)=S(\sigma_0)<+\infty,
\end{equation*}
where $\,S$ is the extension of the von Neumann entropy to the cone $\T_+(\H)$ defined in
(\ref{S-ext}).\smallskip

The approximation technique based on the generalized quantum Dini lemma allows us to prove the relative entropy version of the above result.\smallskip

\textbf{Proposition 3.}  \emph{Let $\,\{\rho_n\}$, $\{\sigma_n\}$ and $\,\{\omega_n\}$  be sequences of operators in $\,\T_+(\H)$ converging, respectively,
to operators  $\rho_0$, $\sigma_0$, and $\,\omega_0$. If
\begin{equation}\label{2-rel}
\!\!\lim_{n\to+\infty}D(\rho_n\|\shs\omega_n)=D(\rho_0\|\shs\omega_0)<+\infty\quad \textit{and}
\quad\lim_{n\to+\infty}D(\sigma_n\|\shs\omega_n)=D(\sigma_0\|\shs\omega_0)<+\infty\!
\end{equation}
then}
\begin{equation}\label{d-rel+}
\lim_{n\to+\infty}D(\rho_n+\sigma_n\|\shs \omega_n)=D(\rho_0+\sigma_0\|\shs \omega_0)<+\infty.
\end{equation}

\emph{Proof.} At the first step we assume that the sequences $\,\{\rho_n\}$ and $\{\sigma_n\}$ consist of states and converge, respectively,
to states  $\rho_0$ and $\sigma_0$. We also assume that $\,\omega_0$ is a nonzero operator. We will prove, by using
Theorem \ref{convex-m} in Section 4.2, that the limit relations in (\ref{2-rel})
imply that
\begin{equation}\label{d-rel++}
\lim_{n\to+\infty}D(p_n\rho_n+\bar{p}_n\sigma_n\|\shs \omega_n)=D(p_0\rho_0+\bar{p}_0\sigma_0\|\shs \omega_0)<+\infty,
\end{equation}
for any sequence $\,\{p_n\}\subset[0,1]$  converging to  $p_0\in[0,1]$, where $\,\bar{p}_n=1-p_n$.

Let $\Psi_m^{\rho}(\rho_n)=P_{\widehat{m}(\rho_0)}^{\rho_n}\rho_n$ and $\Psi_m^{\sigma}(\sigma_n)=P^{\sigma_n}_{\widehat{m}(\sigma_0)}\sigma_n$,
where $P_{\widehat{m}(\rho_0)}^{\rho_n}$ and $P_{\widehat{m}(\sigma_0)}^{\sigma_n}$
are the spectral projectors of $\rho_n$ and $\sigma_n$ defined according to the rule described at the begin of the proof of Theorem \ref{DCT-II}, where it is mentioned that
these families satisfy all the conditions stated at the begin of Section 3 with $\A_\rho=\{\rho_n\}_{n\geq0}$ and $\A_\sigma=\{\sigma_n\}_{n\geq0}$ correspondingly.
In this case $M_{\rho_0}=[m_{\rho},+\infty)\cap\mathbb{N}\,$ and $M_{\sigma_0}=[m_{\sigma},+\infty)\cap\mathbb{N}\,$ for some $m_\rho,m_\sigma>0$.
Let $\mu_n^m=\Tr\Psi_m^{\rho}(\rho_n)$ and $\nu_n^m=\Tr \Psi_m^{\sigma}(\sigma_n)$. By Lemma \ref{DTL-L} in Section 3 we have
\begin{equation}\label{mu-nu+}
\lim_{m\to+\infty}\inf_{n\geq0}\mu^m_n=\lim_{m\to+\infty}\inf_{n\geq0}\nu^m_n=1.
\end{equation}
Denote by $\rho_n^m$ and $\sigma_n^m$ the states $(\mu_n^m)^{-1}\Psi_m^{\rho}(\rho_n)$ and $(\nu_n^m)^{-1}\Psi_m^{\sigma}(\sigma_n)$ for all $n\geq0$
and all $m$ large enough.

Let $f_n(\rho)=D(\rho\shs\|\shs\omega_n)$ for all $n\geq0$. The sequence $\{f_n\}$ consists of lower semicontinuous convex  nonnegative functions satisfying inequality \eqref{LAA-1} with $a_f=h_2$ \cite[Proposition 5.24]{O&P}. This sequence satisfies condition (\ref{f-cond}) due to the lower semicontinuity of the function
$(\rho,\omega)\mapsto D(\rho\shs\|\shs\omega)$. So, to prove (\ref{d-rel++}), it suffices, by Theorem \ref{convex-m} in Section 3,
to verify the validity of condition \eqref{sum-rel}.

It follows from \eqref{mu-nu+} that $\rho_n^m\leq2\rho_n$ and $\sigma_n^m\leq2\sigma_n$ for all $n\geq0$ and all sufficiently large $m$.
So, by Proposition 2 the limit relations in \eqref{2-rel}  imply that
\begin{equation}\label{2-rel+}
\lim_{n\to+\infty}f_n(\rho^m_n)=f_0(\rho^m_0)<+\infty\quad \textup{and}
\quad \lim_{n\to+\infty}f_n(\sigma^m_n)=f_0(\sigma^m_0)<+\infty
\end{equation}
for all sufficiently large $m$ in $M_{\rho_0}\cap M_{\sigma_0}$.
Since
\begin{equation}\label{2-rel+-}
\lim_{n\to+\infty}S(\rho^m_n)=S(\rho^m_0)<+\infty\quad \textup{and}
\quad \lim_{n\to+\infty}S(\sigma^m_n)=S(\sigma^m_0)<+\infty
\end{equation}
for all $m$ in $M_{\rho_0}\cap M_{\sigma_0}$, it follows from the limit relations in \eqref{2-rel+} and representation \eqref{re-exp} that
\begin{equation}\label{2-rel++}
\lim_{n\to+\infty}\Tr H_n\rho^m_n=\Tr H_0\rho^m_0<+\infty\quad \textup{and}
\quad \lim_{n\to+\infty}\Tr H_n\sigma^m_n=\Tr H_0\sigma^m_0<+\infty,
\end{equation}
where $H_n=-\ln\omega_n$ for all $n\geq0$, for all sufficiently large $m$ in $M_{\rho_0}\cap M_{\sigma_0}$. By the remark after  Corollary 4 in \cite{SSP} and because the function $\varrho\mapsto \Tr H_n \varrho$ is affine for each $n$, the limit relations in \eqref{2-rel+-} and \eqref{2-rel++}
imply that
\begin{equation*}%\label{2-rel+}
\lim_{n\to+\infty}S(\theta_n^m)=S(\theta_0^m)<+\infty\quad \textup{and}
\quad \lim_{n\to+\infty}\Tr H_n\theta_n^m=\Tr H_0\theta_0^m<+\infty,
\end{equation*}
where $\theta^m_n=p_n\rho^m_n+\bar{p}_n\sigma^m_n$ for all $n\geq0$, and hence
\begin{equation*}\label{2-rel+++}
\lim_{n\to+\infty}f_n(p_n\rho^m_n+\bar{p}_n\sigma^m_n)=f_0(p_0\rho^m_0+\bar{p}_0\sigma^m_0)<+\infty
\end{equation*}
for all sufficiently large $m$ in $M_{\rho_0}\cap M_{\sigma_0}$ and for any sequence $\,\{p_n\}$ in $[0,1]$  converging to a number $p_0$. This means the validity of condition \eqref{sum-rel}. The implication (\ref{2-rel})$\Rightarrow$(\ref{d-rel++}) is proved.

By using the result of  the first step and identity (\ref{D-mul}) it is easy to establish the implication (\ref{2-rel})$\Rightarrow$(\ref{d-rel+}) in the case when all the operators $\rho_0$, $\sigma_0$ and $\omega_0$ are nonzero.

To prove that (\ref{2-rel}) implies (\ref{d-rel+}) in the general case note that
\begin{equation}\label{D-12+}
 D(\rho_n+\sigma_n\|\shs\omega_n)\leq D(\rho_n\|\shs\omega_n)+D(\sigma_n\|\shs\omega_n)+H(\{\Tr\rho_n,\Tr\sigma_n\})-\Tr \omega_n
\end{equation}
for all $n\geq0$ by inequality (\ref{re-2-ineq-conv}). If $\rho_0$ and $\omega_0$ are arbitrary operators (in particular, $\rho_0=\omega_0=0$) and $\sigma_0=0$  then  inequality (\ref{D-12+}) and the limit relations in (\ref{2-rel}) show that
$$
\limsup_{n\to+\infty}D(\rho_n+\sigma_n\|\shs\omega_n)\leq D(\rho_0\|\shs\omega_0),
$$
since $H(\{\Tr\rho_n,\Tr\sigma_n\})$ tends to zero as $n\to+\infty$.
This inequality and the lower semicontinuity of the relative entropy imply the validity of (\ref{d-rel+}).
$\square$\smallskip

By using Propositions 2 and 3 it is easy to obtain the following

\begin{corollary}\label{re-br++cor}   \emph{Let $\,\{\rho_n\}$, $\{\sigma_n\}$, $\{\omega_n\}$ and $\,\{\vartheta_n\}$  be sequences of operators in $\,\T_+(\H)$ converging, respectively,
to operators  $\rho_0$, $\sigma_0$, $\omega_0$ and $\,\vartheta_0$. If
\begin{equation*}%\label{2-rel}
\!\!\lim_{n\to+\infty}D(\rho_n\|\shs\omega_n)=D(\rho_0\|\shs\omega_0)<+\infty\quad \textit{and}
\quad\lim_{n\to+\infty}D(\sigma_n\|\shs\vartheta_n)=D(\sigma_0\|\shs\vartheta_0)<+\infty\!
\end{equation*}
then}
\begin{equation*}%\label{d-rel+}
\lim_{n\to+\infty}D(\rho_n+\sigma_n\|\shs \omega_n+\vartheta_n)=D(\rho_0+\sigma_0\|\shs \omega_0+\vartheta_0)<+\infty.
\end{equation*}
\end{corollary}
The claim of Corollary \ref{re-br++cor} can be interpreted as preserving local continuity of the quantum relative entropy
under summation.

\subsection{The mutual information of a quantum channel}

The mutual information $I(\Phi,\rho)$ of a  quantum channel $\Phi:A\to B$ at a state $\rho$ in $\S(\H_A)$
can be defined as
\begin{equation}\label{MI-Ch-def}
I(\Phi,\rho)=I(B\!:\!R)_{\Phi\otimes\id_R(\hat{\rho})},
\end{equation}
where $\hat{\rho}$ is pure state in $\S(\H_A\otimes\H_R)$ such that $\Tr_R\hat{\rho}=\rho$ \cite{H-SCI,Wilde}.
This quantity is an important characteristic of a channel related to its  classical entanglement-assisted capacity \cite{BSST,H-c-ch}.

The function $\rho\mapsto I(\Phi,\rho)$ is lower semicontinuous nonnegative and concave on $\S(\H_A)$.
It satisfies inequality
(\ref{LAA-2}) with $b_f(p)=2h_2(p)$, where $h_2$ is the binary entropy \cite[Section 4.3]{MCB}.
Theorems \ref{DCT-II} and \ref{convex-m}  allow us to prove the following\smallskip

\textbf{Proposition 4.} \emph{Let $\{\Phi_n\}$ be a sequence of quantum channels from $A$ to $B$ strongly converging to a
channel $\Phi_0$.\footnote{It means that $\Phi_n(\rho)$ tends to $\Phi_0(\rho)$ for any state $\rho$ in $\S(\H_A)$ \cite{CSR,Wilde+}.} Let $\,\{\rho_n\}$ and $\,\{\sigma_n\}$ be sequences of states in $\S(\H_A)$ converging, respectively,
to states  $\rho_0$ and $\sigma_0$.}

\noindent A) \emph{If $c\rho_n\leq \sigma_n$ for all $n\geq0$ and some $c>0$ and there exists
\begin{equation}\label{MI-sigma}
\lim_{n\to+\infty}I(\Phi_n,\sigma_n)=I(\Phi_0,\sigma_0)<+\infty
\end{equation}
then}
\begin{equation}\label{MI-rho}
\lim_{n\to+\infty}I(\Phi_n,\rho_n)=I(\Phi_0,\rho_0)<+\infty.
\end{equation}

\noindent B) \emph{The limit relation (\ref{MI-rho}) holds if and only if there is
a double sequence $\{P^n_m\}_{n\geq0,m>m_0}$ of finite rank projectors
consistent  with the sequence $\{\rho_n\}$ (i.e. satisfying the conditions in (\ref{P-prop}))
such that
\begin{equation}\label{IP-cont+}
\lim_{m\to+\infty}\sup_{n\geq n_0} \tilde{I}(\Phi_n,\bar{P}^n_m\rho_n\bar{P}^n_m)=0
\end{equation}
for some $n_0>0$, where $\bar{P}^n_m=I_{\H}-P^n_m$ and $\,\tilde{I}(\Phi_n,\varrho)$ is the homogeneous
extension of the function $\varrho \mapsto I(\Phi_n,\varrho)$ to the cone $\T_+(\H_A)$ defined according to the rule (\ref{h-ext}).}\smallskip

\noindent C) \emph{If both limit relations (\ref{MI-sigma}) and (\ref{MI-rho}) hold for the sequences $\,\{\rho_n\}$ and $\,\{\sigma_n\}$ then
\begin{equation*}
\lim_{n\to+\infty}I(\Phi_n,p_n\rho_n+\bar{p}_n\sigma_n)=I(\Phi_0,p_0\rho_0+\bar{p}_0\sigma_0)<+\infty,\quad \bar{p}_n=1-p_n,
\end{equation*}
for any sequence $\,\{p_n\}\subset[0,1]$  converging to  $p_0$.}\smallskip

Claim A of Proposition 4 can be treated as a strengthened form of the dominated convergence theorem for the von Neumann entropy
mentioned at the begin of Section 4.2, since  $I(\Phi,\rho)=2S(\rho)$ if $\Phi$ is the identity channel.\smallskip

\emph{Proof.} Consider the sequence of lower semicontinuous nonnegative concave functions $f_n(\rho)=I(\Phi_n,\rho)$ on $\S(\H_A)$
satisfying inequality (\ref{LAA-2}) with $b_f(p)=2h_2(p)$.\smallskip

A) This claim is proved by applying  Theorem \ref{DCT-II}B in Section 4.2 to the sequence $\{f_n\}$. Indeed, condition (\ref{n-l-s-c}) holds
for this sequence due to the lower semicontinuity of the function $\varrho\mapsto I(B\!:\!R)_{\varrho}$, since the strong convergence
of the sequence $\{\Phi_n\}$ to the channel $\Phi_0$ implies that (cf.\cite{Wilde+})
\begin{equation}\label{st-conv}
\lim_{n\to+\infty}\Phi_n\otimes\id_{R}(\hat{\rho}_n)=\Phi_0\otimes\id_{R}(\hat{\rho}_0)
\end{equation}
for any sequence $\{\hat{\rho}_n\}$ of pure states in $\S(\H_A\otimes\H_R)$ converging to a pure state
$\hat{\rho}_0$ such that $\Tr_{R}\hat{\rho}_n=\rho_n$ for all $n\geq0$.
Thus, it suffices to show that
\begin{equation}\label{b-cond-II+}
\lim_{n\to+\infty} f_n(\rho_n)=f_0(\rho_0)<+\infty
\end{equation}
for any sequence $\{\rho_n\}\subset\S(\H_A)$ converging to a state $\rho_0$ such that $\sup_n\rank \rho_n<+\infty$,
since this obviously implies the validity of condition (\ref{gen-con-II}) with any $c>0$.

To prove (\ref{b-cond-II+}) note that the assumption $\sup_n\rank \rho_n\leq d<+\infty$ implies existence
of a sequence $\{\hat{\rho}_n\}$ of pure states in $\S(\H_A\otimes\H_R)$ converging to a pure state
$\hat{\rho}_0$ such that $\Tr_{R}\hat{\rho}_n=\rho_n$ for all $n\geq0$, where $\H_R$ is a $d$-dimensional Hilbert space.
By using Remark 1 in Section 2 and taking (\ref{st-conv}) into account we obtain (\ref{b-cond-II+}).\smallskip

B) The "if" part of  claim B is proved by applying  Corollary \ref{DCT-II-c} in Section 4 to the sequence $\{f_n\}$. It suffices to
use the arguments from the proof of claim A justifying the applicability of  Theorem \ref{DCT-II} to the sequence $\{f_n\}$  and to show that
\begin{equation*}%\label{IP-cont}
\lim_{n\to+\infty} \tilde{I}(\Phi_n,P^n_m\rho_nP^n_m)=\tilde{I}(\Phi_0,P^0_m\rho_0P^0_m)<+\infty\quad \forall m\geq m_0,
\end{equation*}
what can be done by noting that $\rank P^n_m\rho_nP^n_m\leq m$ and by using (\ref{b-cond-II+}).

To prove the "only if" part of  claim B assume that (\ref{MI-rho}) holds and $I(\Phi_n,\rho_n)<+\infty$ for all $n>n_0$.
Take a double sequence $\{P^n_m\}_{n\geq0,m\geq m_0}$ ($m_0\in\mathbb{N}$) of finite rank projectors consistent
with the sequence $\{\rho_n\}$ such that $P^n_m\rho_n=\rho_n P^n_m$  for all $n\geq0$, $m\geq m_0$, which exists by Lemma \ref{imp-l} below.
By using the concavity of the function $f_n(\varrho)=I(\Phi_n,\varrho)$ it is easy to show that
\begin{equation}\label{c-in-1}
\tilde{I}(\Phi_n,P^n_{m}\rho_n)\leq\tilde{I}(\Phi_n,P^n_{m+1}\rho_n)\leq I(\Phi_n,\rho_n)
\end{equation}
and that
\begin{equation}\label{c-in-2}
\tilde{I}(\Phi_n,P^n_m\rho_n)+\tilde{I}(\Phi_n,\bar{P}^n_m\rho_n)\leq I(\Phi_n,\rho_n)
\end{equation}
for all $m\geq m_0$ and $n\geq0$. By using the lower semicontinuity of the function $f_n(\varrho)=I(\Phi_n,\varrho)$
and the second inequality in (\ref{c-in-1}) we obtain
$$
\lim_{m\to+\infty}\tilde{I}(\Phi_n,P^n_{m}\rho_n)=I(\Phi_n,\rho_n)<+\infty\quad \forall n\geq n_0,\;n=0.
$$
By using the lower semicontinuity of the quantum mutual information it is easy to show that
$$
\liminf_{n\to+\infty}\tilde{I}(\Phi_n,P^n_{m}\rho_n)\geq \tilde{I}(\Phi_0,P^0_{m}\rho_0)\quad \forall m\geq m_0.
$$
The last  limit relations and the first inequality in (\ref{c-in-1}) imply, by Dini's lemma, that
$\tilde{I}(\Phi_n,P^n_{m}\rho_n)$ tends to $I(\Phi_n,\rho_n)$ as $m\to+\infty$ uniformly on $n\geq n_0$.
Thus, inequality (\ref{c-in-2}) shows that (\ref{IP-cont+}) holds.\smallskip

C) Let $\Psi_m^{\rho}(\rho_n)=P_{\widehat{m}(\rho_0)}^{\rho_n}\rho_n$ and $\Psi_m^{\sigma}(\sigma_n)=P^{\sigma_n}_{\widehat{m}(\sigma_0)}\sigma_n$,
where $P_{\widehat{m}(\rho_0)}^{\rho_n}$ and $P_{\widehat{m}(\sigma_0)}^{\sigma_n}$
are the spectral projectors of $\rho_n$ and $\sigma_n$ defined according to the rule described at the begin of the proof of Theorem \ref{DCT-II} in Section 4.2.
Then by using (\ref{b-cond-II+}) it is easy to show that
all the conditions of Theorem \ref{convex-m} in Section 4.2 are satisfied for the sequence $\{f_n\}$. $\Box$\smallskip

\begin{lemma}\label{imp-l} \emph{For any sequence $\{\rho_n\}\subset\T_+(\H)\setminus\{0\}$ converging to a nonzero operator $\rho_0$
there exists a double sequence $\{P^n_m\}_{n\geq0,m\geq m_0}$ ($m_0\in\mathbb{N}$) of finite rank projectors consistent
with the sequence $\{\rho_n\}$ (i.e. satisfying the conditions in (\ref{P-prop})) such that $P^n_m\rho_n=\rho_n P^n_m$  for all $n\geq0$, $m\geq m_0$.}
\end{lemma}\smallskip

\emph{Proof.} Assume first that all the operators $\rho_n$, $n\geq0$, have infinite rank.
In this case we set $P^n_m=P_{\widehat{m}(\rho_0)}^{\rho_n}$ for all $m\geq m_0$, where $P_{\widehat{m}(\rho_0)}^{\rho_n}$ is the spectral projector of $\rho_n$  defined according to the rule
described at the begin of the proof of Theorem \ref{DCT-II} in Section 4.2 and $m_0$ is the
multiplicity of the maximal eigenvalue of $\rho_0$ ($\widehat{m}(\rho_0)$ is well defined for all $m\geq m_0$).

By using the arguments presented at the end of Section 4.2.1 in \cite{QC}
(based on Theorem VIII.23 in \cite{R&S} and the Mirsky inequality) it is easy to show that
\begin{equation*}%\label{P-lim}
   P_{\widehat{m}(\rho_0)}^{\rho_0}=\lim_{n\to+\infty} P_{\widehat{m}(\rho_0)}^{\rho_n} \quad \textrm{in the operator norm for all}\;\; m\geq m_0,
\end{equation*}
which implies the last property in (\ref{P-prop}) for the sequence $\{P^n_m\}_{n\geq0,m\geq m_0}$ constructed in this way. The validity of the others
properties in (\ref{P-prop}) directly follows from the construction.

Assume now that the operators $\rho_n$, $n\geq0$, have arbitrary rank. Take an infinite rank quantum
state $\sigma$ on an auxiliary Hilbert space $\H'$ such that
\begin{equation}\label{spect}
  \|\sigma\|<\|\rho_n\|\quad \textrm{and}\quad \mathrm{spect}(\sigma)\cap\mathrm{spect}(\rho_n)=\{0\}\quad \forall n\geq0,
\end{equation}
where $\mathrm{spect}$ denotes the spectrum. Then the sequence of infinite rank operators $\tilde{\rho}_n=\rho_n\oplus\sigma$ in $\T_+(\H\oplus\H')$
converges to the infinite rank operator $\tilde{\rho}_0=\rho_0\oplus\sigma$. Let $\{\tilde{P}^n_m\}_{n\geq0,m\geq m_0}$
be the double sequence of finite rank projectors constructed for the sequence $\{\tilde{\rho}_n\}$ by the rule
described at the begin of this proof. The second condition in (\ref{spect}) guarantees that all the projectors $\tilde{P}^n_m$
have the form $P^n_m\oplus Q^n_m$, where $P^n_m$ and $Q^n_m$ are spectral projectors of the operators $\rho_n$ and $\sigma$ correspondingly.
The first condition in (\ref{spect}) implies that $P^n_m>0$ for all $n\geq0$ and $m\geq m_0$. It is easy to see that $\{P^n_m\}_{n\geq0,m\geq m_0}$
is a double sequence of finite rank projectors having the required properties. $\Box$\smallskip

The convergence criterion for the function $\,(\Phi,\rho)\mapsto I(\Phi,\rho)\,$ given by Proposition 4B is quite powerful.
It is easy to see that it gives another ways to prove claims A and C of Proposition 4 by using only the LAA property of the function $\,\rho\mapsto I(\Phi,\rho)$.
This criterion implies the following result that can be treated as preserving convergence of the mutual information
under concatenation.\smallskip

\begin{corollary}\label{conc-cont}
\emph{Let $\,\{\rho_n\}$ be a sequences of states in $\S(\H_A)$ converging
to a state  $\rho_0$.}\smallskip

\emph{If $\{\Phi_n\}$ is a sequence of quantum channels from $A$ to $B$ strongly converging to a
channel $\Phi_0$ such that limit relation (\ref{MI-rho}) holds then
\begin{equation*}%\label{MI-rho}
\lim_{n\to+\infty}I(\Psi_n\circ\Phi_n,\rho_n)=I(\Psi_0\circ\Phi_0,\rho_0)<+\infty
\end{equation*}
for arbitrary sequence $\{\Psi_n\}$ of quantum channels from $B$ to $C$ strongly converging to a
channel $\Psi_0$.}
\end{corollary}\smallskip

\emph{Proof.} It suffices to note that the chain rule for the mutual information (cf. \cite{H-SCI,Wilde}) implies that
$\tilde{I}(\Psi_n\circ\Phi_n,\sigma)\leq \tilde{I}(\Phi_n,\sigma)\,$
for all $n$ and any operator $\sigma\in\T_+(\H_A)$ and to apply  Proposition 4B. $\Box$ \smallskip

In the case $\Phi_n=\id_A$ for all $n\geq0$
(in which $I(\Phi_n,\rho)=2S(\rho_n)$) Proposition 4B  gives a convergence criterion for the von Neumann entropy, i.e. criterion for validity of the limit relation
\begin{equation}\label{S-conv}
\lim_{n\to+\infty}S(\rho_n)=S(\rho_0)<+\infty
\end{equation}
for a sequence $\{\rho_n\}\subset\S(\H_A)$ converging to a state $\rho_0$.  In fact, in this case the "only if" part of the criterion can be made more stronger as shown by claim B of the following\smallskip

\begin{corollary}\label{S-conv-crit} \emph{Let $\,\{\rho_n\}$ be a sequence of states in $\S(\H)$ converging to a
state $\rho_0$.}\smallskip

A) \emph{The limit relation (\ref{S-conv}) holds if there is
a double sequence $\{P^n_m\}_{n\geq0,m>m_0}$ of finite rank projectors
consistent  with the sequence $\{\rho_n\}$ (i.e. satisfying the conditions in (\ref{P-prop}))
such that
\begin{equation}\label{S-cont+}
\lim_{m\to+\infty}\sup_{n\geq n_0} S(\bar{P}^n_m\rho_n\bar{P}^n_m)=0
\end{equation}
for some $n_0>0$, where $\bar{P}^n_m=I_{\H}-P^n_m$ and $\,S$ is the homogeneous
extension of the von Neumann entropy to the positive cone $\T_+(\H)$ defined in
(\ref{S-ext}).}\smallskip

B) \emph{If the limit relation (\ref{S-conv}) is valid then (\ref{S-cont+}) holds for
\textbf{any} double sequence $\{P^n_m\}_{n\geq0,m>m_0}$ of finite rank projectors
consistent  with the sequence $\{\rho_n\}$.}
\end{corollary}\medskip

\emph{Proof.} Claim A is a direct corollary of Proposition 4B with $\Phi_n=\id_A$ for all $n\geq0$.\smallskip

Claim B is proved by using the arguments from the proof of the "only if" part of Proposition 4B
and by noting that for \emph{any} double sequence $\{P^n_m\}_{n\geq0,m>m_0}$ of finite rank projectors
consistent  with the sequence $\{\rho_n\}$ the following two statements are valid:
\begin{itemize}
  \item $S(P^n_m\rho_nP^n_m)$ monotonously increases to $S(\rho_n)$ as $m\rightarrow+\infty$ for all $n\geq 0$ by Lemma 4 in \cite{L-2};
  \item $S(P^n_m\rho_nP^n_m)+S(\bar{P}^n_m\rho_n\bar{P}^n_m)\leq S(\rho_n)$ for all $n\geq 0$ and $m\geq m_0$  by the Lindblad-Ozawa inequality for the POVM $\{P^n_m, \bar{P}^n_m\}$ \cite{L-ER}. $\Box$
\end{itemize}

Corollary \ref{S-conv-crit} gives a simple way to prove many of the convergence conditions
for the von Neumann entropy obtained in \cite[Appendix]{Ruskai},\cite{SSP,W}.\smallskip

\textbf{Example 2.} By using Proposition 4B one can  show
that the limit relation (\ref{MI-rho}) holds for a sequence $\,\{\rho_n\}$  of states in $\S(\H_A)$ converging to a state  $\rho_0$ and
a  sequence $\{\Phi_n\}$ of channels strongly converging  to a channel $\Phi_0$ provided that
\begin{equation}\label{2-cond}
\!\textrm{either}\;\; \lim_{n\to+\infty}S(\rho_n)=S(\rho_0)<+\infty\;\;\; \textup{or}
\;\; \lim_{n\to+\infty}S(\Phi_n(\rho_n))=S(\Phi_0(\rho_0))<+\infty.\!
\end{equation}
Indeed, the upper bound (\ref{MI-UB}) for the quantum mutual information shows that
\begin{equation*}%\label{I-UB}
  \tilde{I}(\Phi,\rho)\leq 2\min\{S(\rho),S(\Phi(\rho))\}
\end{equation*}
for any operator $\rho$ in $\T_+(\H_A)$ and any channel $\Phi:A\to B$.
So, if the first condition in (\ref{2-cond}) is valid then Corollary \ref{S-conv-crit}B
implies existence of a double sequence $\{P^n_m\}_{n\geq0,m>m_0}$ of finite rank projectors
consistent  with the sequence $\{\rho_n\}$ such that (\ref{IP-cont+}) holds.
If the second condition in (\ref{2-cond}) is valid then the same conclusion
can be done by using Lemma \ref{five} below.\smallskip

The implication (\ref{2-cond})$\Rightarrow$(\ref{MI-rho}) can be also derived from Theorem 1B in Section 4.1 by modifying
the arguments from  Example 1 in Section 4.1 or by using Proposition 18 in \cite[Section 5.2.3]{QC}.\smallskip

\begin{lemma}\label{five} \emph{If $\{\Phi_n\}$ is a sequence of quantum channels from $A$ to $B$ strongly converging to a
channel $\Phi_0$ and $\,\{\rho_n\}$  is a sequence of states in $\S(\H_A)$ converging to a
state $\rho_0$ such that the second condition in (\ref{2-cond}) holds then
there is a double sequence $\{P^n_m\}_{n\geq0,m>m_0}$ of finite rank projectors
consistent  with the sequence $\{\rho_n\}$ such that
\begin{equation*}%\label{S-cont+}
\lim_{m\to+\infty}\sup_{n\geq n_0} S(\Phi_n(\bar{P}^n_m\rho_n\bar{P}^n_m))=0
\end{equation*}
for some $n_0>0$, where $\bar{P}^n_m=I_{\H}-P^n_m$ and $\,S$ is the homogeneous
extension of the von Neumann entropy to the positive cone $\T_+(\H)$ defined in
(\ref{S-ext}).}
\end{lemma}\smallskip

\emph{Proof.} It suffices to repeat the arguments from the proof of the "only if" part of Proposition 4B
with $\tilde{I}(\Phi_n,(\cdot))$ replaced by $S(\Phi_n(\cdot))$. $\Box$

\section*{Appendix}

We assume below that the value of $\Tr H \rho$ (finite or infinite) for a  positive (semi-definite) operator $H$ on a Hilbert space $\H$ and any positive operator $\rho$ in $\T(\H)$ is defined according to the rule (\ref{H-as}).\pagebreak

\textbf{Proposition 5.}  \emph{Let $\{\rho^1_n\}$, $\{\rho^2_n\}$, $\{\sigma^1_n\}$ and $\{\sigma^2_n\}$ be sequences of  operators in $\T_+(\H)$ converging, respectively,
to nonzero operators $\rho^1_0$, $\rho^2_0$, $\sigma^1_0$ and $\sigma^2_0$ such that $\,\rho^1_n\geq \rho^2_n$ and $\,\sigma^1_n\leq \sigma^2_n$ for all $\,n\geq0$. If
\begin{equation}\label{c-11}
A_{1}\doteq\limsup_{n\to+\infty} \Tr \rho^1_n(-\ln \sigma_n^1)<+\infty\quad and \quad A_{1}-\Tr \rho^1_0(-\ln \sigma_0^1)=\Delta
\end{equation}
then}
$$
A_{2}\doteq\limsup_{n\to+\infty} \Tr \rho^2_n(-\ln \sigma_n^2)<+\infty\quad and \quad A_{2}-\Tr \rho^2_0(-\ln \sigma_0^2)\leq\Delta.
$$

\textbf{Remark 5.} The lower semicontinuity of the von Neumann entropy and the relative entropy implies, due to representation \eqref{re-exp}, that the function $\,(\rho, \sigma)\mapsto \Tr \rho(-\ln \sigma)\,$
is lower semicontinuous on $\,\T_+(\H)\times \T_+(\H)$. Hence, the quantities  $\,A_{i}-\Tr \rho^i_0(-\ln \sigma_0^i)$, $i=1,2$, in
Proposition 5 are nonnegative.\smallskip

\emph{Proof.} By the condition \eqref{c-11} we may assume that $\Tr \rho^1_n(-\ln \sigma_n^1)<+\infty$ for all $n\geq 0$ and hence
\begin{equation}\label{supp}
  \supp\rho_n^i \subseteq \supp\sigma_n^i \quad \forall n\geq0,\; i=1,2.
\end{equation}

Let $a_{n}^{i}\doteq\Tr \rho^i_n(-\ln\sigma_n^i)$, $n\geq0$, $i=1,2$. For any natural $k$ introduce the quantities
$$
a_{k,n}^{i}=-\Tr\rho^i_n\ln (\sigma_n^i+k^{-1}I_{\H}),\quad n\geq0,\; i=1,2.
$$
By Theorems VIII.18 and VIII.20 in \cite{R&S} for each $i$ and $k$ the sequence of bounded operators
$\ln (\sigma_n^i+k^{-1}I_{\H})$ tends to the bounded operator
$\ln (\sigma_0^i+k^{-1}I_{\H})$ as $n\to+\infty$ in the operator norm. It follows that
\begin{equation}\label{k-lim-rel}
\lim_{n\to+\infty} a_{k,n}^{i}=a_{k,0}^{i}<+\infty,\quad \forall k\in \mathbb{N},\; i=1,2.
\end{equation}
By using the operator monotonicity of the logarithm  it is easy to show that
\begin{equation}\label{non-dec}
\sup_k a_{k,n}^{i}=a_{n}^{i}, \quad \forall n\geq0,\; i=1,2.
\end{equation}

Thus, for any $\varepsilon>0$  there is $k_{\varepsilon}$ such that
$a_{0}^{1}-a_{k_{\varepsilon},0}^{1}<\varepsilon$. Condition \eqref{c-11} and
relation \eqref{k-lim-rel} with $i=1$ imply that there is $n_{\varepsilon}$ such that  $a_{n}^{1}-a_{0}^{1}\leq\Delta+\varepsilon$ and
$a_{0,k_{\varepsilon}}^{1}-a_{n,k_{\varepsilon}}^{1}\leq \varepsilon$ for all $n\geq n_{\varepsilon}$. It follows that
\begin{equation}\label{uniform-conv}
a_{n}^{1}-a_{k_{\varepsilon},n}^{1}=(a_{n}^{1}-a_{0}^{1})+(a_{0}^{1}-a_{k_{\varepsilon},0}^{1})+(a_{k_{\varepsilon},0}^{1}-a_{k_{\varepsilon},n}^{1})
<\Delta+3\varepsilon \quad \forall n\geq n_{\varepsilon}.
\end{equation}
By using expression \eqref{Tr-exp} and taking \eqref{supp} into account it is easy to show that
$$
a_{n}^{i}-a_{k,n}^{i}=\Tr\rho^i_n \ln (I_{\H}+k^{-1}[\sigma_n^i]^{-1}),\quad\forall k,\forall n\geq0,\; i=1,2,
$$
where $[\sigma_n^i]^{-1}$ is the Moore-Penrose inverse of $\sigma_n^i$. So, since $\rho_n^1\geq\rho_n^2$ and $\sigma_n^1\leq\sigma_n^2$, Lemmas \ref{log-m} and \ref{sqrt} below show that
$$
0\leq a_{n}^{2}-a_{k,n}^{2}\leq a_{n}^{1}-a_{k,n}^{1},\quad \forall k, \forall n\geq0.
$$
Thus, it follows from \eqref{k-lim-rel} with $i=2$, \eqref{non-dec} and \eqref{uniform-conv} that
$$
a_{n}^{2}-a_{0}^{2}\leq a_{n}^{2}-a_{0,k}^{2}=(a_{n}^{2}-a_{n,k_{\varepsilon}}^{2})+(a_{n,k_{\varepsilon}}^{2}-a_{0,k_{\varepsilon}}^{2})\leq \Delta+4\varepsilon
$$
for all sufficiently large $n$. This implies the assertion of the proposition.
$\square$\smallskip

\begin{lemma}\label{log-m}  \emph{Let $\sigma_1$ and $\sigma_2$  be nonzero positive  operators in $\T(\H)$
such that $\sigma_1\leq\sigma_2$ and $c>0$ be arbitrary. Then   $H_1=\ln(I_{\H}+c\sigma^{-1}_1)$ and $H_2=\ln(I_{\H}+c\sigma^{-1}_2)$ are positive semidefinite  operators on $\H$
such that
$$
\D(\sqrt{-\ln \sigma_1})\subseteq\D(\sqrt{H_1})\cap\D(\sqrt{H_2})\quad and \quad \langle\varphi|H_1|\varphi\rangle\geq\langle\varphi|H_2|\varphi\rangle
$$
for any $\varphi\in\D(\sqrt{-\ln \sigma_1})$, where $\langle\varphi|H_i|\varphi\rangle\doteq\|\sqrt{H_i}\varphi\|^2$ and $\sigma^{-1}_i$ denotes the Moore-Penrose inverse of $\sigma_i$, $i=1,2$.}\footnote{We denote by
$\D(H)$ the domain of $H$.}
\end{lemma}\smallskip

If the operators $\sigma_1$ and $\sigma_2$ are non-degenerate then Lemma \ref{log-m} can be deduced easily from Corollary 10.12 in \cite{Sch}
and the operator monotonicity of the logarithm.\smallskip

\emph{Proof.} Denote $I_{\H}$ by $I$ for brevity. For a given vector $\varphi\in\D(\sqrt{-\ln\sigma_1}\shs)$ and any natural $k$ and $n$ introduce the quantities
$$
a_{k,n}^i=\langle\varphi|P_n^i\ln (I+c(\sigma_i+k^{-1}I)^{-1})|\varphi\rangle,\qquad a_{k,*}^i=\langle\varphi|\ln (I+c(\sigma_i+k^{-1}I)^{-1})|\varphi\rangle
$$
and
$$
a_{*,n}^i=\langle\varphi|P_n^i\ln (I+c\sigma_i^{-1})|\varphi\rangle,\quad i=1,2,
$$
where $P_n^i$ is the spectral projector of the operator $\sigma_i$ corresponding to its $n$ maximal nonzero eigenvalues (taking multiplicity into account), $i=1,2$ (if $\rank\sigma_i<n$ we assume that $P_n^i$ is the projector on the support $\supp \sigma_i$ of $\sigma_i$).  Since $\,\varphi\in\supp \sigma_1\subseteq\supp \sigma_2\,$ by the condition
$\sigma_1\leq\sigma_2$, we have
\begin{equation}\label{imp-lr}
\lim_{n\rightarrow+\infty}P_n^1|\varphi\rangle=\lim_{n\rightarrow+\infty}P_n^2|\varphi\rangle=|\varphi\rangle.
\end{equation}
It is easy to see that $\varphi\in\D(\sqrt{H_1}\shs)$, which means that $\,\sup_n a_{*,n}^1\,$ is finite. Thus, to prove the assertion of
the lemma it suffices to show that
$$
\sup_n a_{*,n}^2\leq \sup_n a_{*,n}^1.
$$

The positivity of the operator $\,\ln (I+c(\sigma_i+k^{-1}I)^{-1})\,$ and the operator monotonicity of the logarithm imply that
\begin{equation}\label{one-ineq}
a_{k,n}^i\leq a_{k,n+1}^i,\qquad a_{k,n}^i\leq a_{k+1,n}^i,\quad i=1,2,
\end{equation}
for all $n$ and $k$. Since $\sigma_1\leq\sigma_2$, we have $\,(\sigma_1+k^{-1}I)^{-1}\geq(\sigma_2+k^{-1}I)^{-1}$ for any $k$ \cite[Corollary 10.13]{Sch},
So, the operator monotonicity of the logarithm also implies
\begin{equation}\label{two-ineq}
a_{k,*}^1\geq a_{k,*}^2\quad \forall k.
\end{equation}
Since $\,P_n^i\ln (I+c\sigma_i^{-1})\,$ and $\,\ln(I+c(\sigma_i+k^{-1}I)^{-1})\,$ are bounded positive operators, it follows from  \eqref{imp-lr} and \eqref{one-ineq} that
$$
\sup_n a_{k,n}^i=\lim_{n\rightarrow+\infty} a_{k,n}^i=a_{k,*}^i, \qquad \sup_k a_{k,n}^i=\lim_{k\rightarrow+\infty}a_{k,n}^i=a_{*,n}^i
$$
and hence
$$
\sup_n a_{*,n}^i=\sup_n\sup_k a_{k,n}^i=\sup_k\sup_n a_{k,n}^i=\sup_k a_{k,*}^i,\quad i=1,2.
$$
Thus, inequality (\ref{two-ineq}) shows that
$$
\sup_n a_{*,n}^2=\sup_k a_{k,*}^2\leq\sup_k a_{k,*}^1=\sup_n a_{*,n}^1. \;\;\square
$$

\begin{lemma}\label{sqrt}  \emph{Let  $H$ be a positive (semi-definite) operator on a Hilbert space $\H$ and  $\rho=\sum_{i}|\varphi_i\rangle\langle\varphi_i|$ the spectral decomposition of a positive operator $\rho$ in $\T(\H)$.
If $\Tr H \rho$ is finite then all the vectors $\varphi_i$ belong to the domain of $\sqrt{H}$ and}
$$
\Tr H \rho=\sum_{i}\langle\varphi_i|H|\varphi_i\rangle=\sum_{i}\|\sqrt{H}\varphi_i\|^2.
$$
\end{lemma}

\emph{Proof.} Let $P_n$ be the spectral projector of $H$ corresponding to the interval $[0,n]$. Since $P_nH$ is a bounded operator for each $n$, we have
$$
\Tr HP_n \rho=\sum_{i}\|\sqrt{H}P_n\varphi_i\|^2.
$$
Since the sequence $\{\|\sqrt{H}P_n\varphi_i\|^2\}_n$ is nondecreasing for each given $i$, it follows from the assumption  $\,\Tr H\rho=\sup_n\Tr HP_n \rho<+\infty\,$
that $\,a_i=\sup_n\|\sqrt{H}P_n\varphi_i\|^2<+\infty\,$ for all $i$ and  $\Tr H\rho=\sum_i a_i$. The finiteness of $a_i$ means, by the spectral theorem, that $\varphi_i\in\D(\sqrt{H})$ and $a_i=\|\sqrt{H}\varphi_i\|^2$. $\square$

\bigskip

I am grateful to L.Lami whose study  of the relative entropy of resource
motivated this research.
I am also grateful to A.S.Holevo and to the participants of his seminar
"Quantum probability, statistic, information" (the Steklov
Mathematical Institute) for useful discussion.\smallskip

This work is supported by the
Russian Science Foundation (Grant No. 19-11-00086)

%\bibliographystyle{abbrv}

%\bibliography{library}

\end{document}